\documentclass[letter]{emulateapj}

\usepackage{amssymb} \usepackage{lscape} \usepackage{longtable}

\def\lsim{\mathrel{\rlap{\lower4pt\hbox{\hskip1pt$\sim$}}
    \raise1pt\hbox{$<$}}}                
\def\gsim{\mathrel{\rlap{\lower4pt\hbox{\hskip1pt$\sim$}}
    \raise1pt\hbox{$>$}}}                

\shorttitle{Evolution of LBGs between $z\simeq 4-6$}
\submitted{Submitted to ApJ and revised after referee comments}
\begin{document}

\shortauthors{Stark et al.}

\title{The Evolutionary History of Lyman Break Galaxies Between
  Redshift 4 and 6: Observing Successive Generations of Massive
  Galaxies in Formation}


\author {Daniel P. Stark\altaffilmark{1,2},  Richard
  S. Ellis\altaffilmark{1,3}, Andrew Bunker\altaffilmark{3},  Kevin
  Bundy\altaffilmark{4,5},  Tom Targett\altaffilmark{1,6},  Andrew
  Benson\altaffilmark{1}, Mark Lacy\altaffilmark{7} }

\altaffiltext{1}{Department of Astrophysics, California Institute of
  Technology, MS 105-24, Pasadena, CA 91125}
\altaffiltext{2}{Institute of Astronomy, University of Cambridge,
  Madingley  Road, Cambridge CB3 0HA}  \altaffiltext{3}{Department of
  Astrophysics, University of Oxford, OX1 3RH}
\altaffiltext{4}{Hubble Fellow, Department of Astronomy, University of
  California,  Berkeley, CA 94720} \altaffiltext{5}{Reinhardt Fellow,
  Department of Astronomy \& Astrophysics, University of Toronto, 50
  St. George Street, Toronto, ON, M5S 3H4} \altaffiltext{6}{Department
  of Physics and Astronomy, University of British Columbia, 6224
  Agricultural Rd., Vancouver, B.C., V6T 1Z1, Canada.}
\altaffiltext{7}{Spitzer Science Center, California Institute of
  Technology, MC-220-6, 1200 E. California Blvd, Pasadena, CA 91125,
  U.S.A.}

\begin{abstract} 

We present new measurements of the evolution in the Lyman break galaxy
(LBG) population between $z\simeq 4$ and $z\simeq 6$.  By utilizing
the extensive multiwavelength datasets available in the GOODS fields,
we identify 2443 B, 506 V, and 137 $i'$-band dropout galaxies likely
to be at $z\approx 4$, $5$, \& $6$.  For the subset of dropouts for
which reliable {\em Spitzer} IRAC photometry is feasible (roughly 35\%
of the sample), we estimate luminosity-weighted ages and stellar
masses.  With the goal of understanding the duration of typical star
formation episodes in galaxies at $z\gsim 4$, we examine the
distribution of stellar masses and ages as a function of cosmic time.
We find that at a fixed rest-UV luminosity, the average stellar masses
and ages of galaxies do not increase significantly between  $z\simeq
6$ and 4. In order to maintain this near equilibrium in the average
properties of high redshift LBGs, we argue that there must be a steady
flux of young,  newly-luminous objects at each successive redshift.
When considered along  with the short duty cycles inferred from
clustering measurements, these results may suggest that galaxies are
undergoing star formation episodes lasting only several hundred
million years.  In contrast to the unchanging relationship between the
average stellar mass and rest-UV luminosity, we find that the number
density of massive  galaxies increases considerably with time over
$4\lsim z\lsim 6$.  Given this rapid increase of UV luminous massive
galaxies, we explore the possibility that a significant fraction of
massive (10$^{11}$ M$_\odot$) $z\simeq 2-3$ distant red galaxies
(DRGs) were in part assembled in an LBG phase at earlier times.
Integrating the growth in the stellar mass function of actively
forming LBGs over $4\lsim z\lsim 6$ down to $z\simeq 2$, we find that
$z\gsim 3$ LBGs could have contributed significantly to the quiescent
DRG population, indicating that the intense star-forming systems
probed by sub-millimeter observations are not the only route toward
the assembly of DRGs at $z\simeq 2$. 

\end{abstract} 

\section{Introduction}
\label{sec:intro}

The detailed study of various classes of distant galaxies has enabled
great progress in understanding the star formation and mass assembly
history of normal field galaxies (for recent reviews see
\citealt{Hopkins06, Ellis08, Wilkins08}).  Multi-wavelength probes
have been particularly effective in revealing the co-existence of
diverse categories of galaxies with redshifts $z\simeq 2$-3. These
include the  relatively unobscured star-forming `Lyman break' galaxies
(LBGs, e.g., \citealt{Steidel96, Shapley05}), the infrared-selected
massive `distant  red' galaxies (DRGs, e.g., \citealt{Franx03,
  vanDokkum06}) and heavily obscured  sub-mm galaxies which contain
both intensely star-forming and active  components (SMGs,
e.g. \citealt{Smail98,Chapman05}). The collective study  of these
populations has revealed that the redshift range $1<z<3$  is a
formative one when the bulk of the stars in present-day massive
galaxies  was produced \citep{Hopkins06}.

Understanding the inter-relationship between these various sources is
an important  goal and intense efforts are now underway to address
this issue (e.g., \citealt{vanDokkum06,Reddy08}). A relevant aspect of
this discussion concerns the assembly  history of objects observed
during the redshift interval $4<z<6$, corresponding to a period only 1
Gyr earlier.  Such data may provide valuable insight into the
connection between actively star-forming and passive populations as
well as define the mode of star formation in typical massive galaxies.

Over the last five years, deep multiwavelength surveys have resulted
in the discovery of large samples of LBGs at $z\simeq 4-6$
\citep{Bouwens07}.  Despite early controversies
\citep{Bunker04,Stanway03,Giavalisco04b,Beckwith06}, it now seems
clear that the star formation density  declines with redshift beyond
$z\simeq 3$. Recent evidence also suggests  the characteristic
luminosity is also fading \citep{Yoshida06,Bouwens07, McLure08}.
\cite{Bouwens07}  attribute this evolutionary pattern to the simple
hierarchical assembly  of galaxies. Unfortunately, because of the
transient nature of star formation probed by the rest-frame UV
luminosity function, these studies alone provide  only an approximate
measure of the evolutionary processes occurring during $3<z<6$.  Key
to testing the mode of assembly of galaxies in this early period is
additional information on the physical properties, such as the
associated stellar mass and the inferred age of the stellar
populations. The availability of deep IRAC data for many of the HST
fields enables such an approach \citep{Egami05,Eyles05,HYan05}.
 
In this paper, we aim to improve our understanding of galaxy evolution
during the first 2 Gyr by systematically tracking the evolving stellar
content of star-forming galaxies in uniformly-selected samples
spanning three separate redshift intervals between $z\simeq 4$ and
6. By adding the additional physical parameters of stellar mass and
luminosity-weighted ages, we can break degeneracies associated with
studies that rely only on the UV luminosity function.

An important issue is how and when the quiescent subset of DRGs seen
at  $z\simeq 2-3$ \citep{vanDokkum06,Kriek06} assembled their mass.
Contemporary models of galaxy formation suggest that the number
density of massive galaxies increases continuously with cosmic time
over $2\lsim z\lsim 5$ \citep{Bower06}, a picture supported by the
relatively young ages of massive DRGs at $z\simeq 2$ \citep{Kriek06}.
Alternatively, it is potentially feasible that the bulk of the most
massive  systems  at $z\simeq 2$ formed their mass at much earlier
times (e.g., $z\gsim 5$) which in turn would imply little evolution in
the number density of massive  galaxies over 3$\lsim z\lsim 6$.  We
seek to constrain the formation epoch  of quiescent DRGs by studying
the evolving stellar mass function of star-forming galaxies over
$4\lsim z\lsim 6$.  Moreover, we wish to understand whether most DRGs
were assembled entirely in intense, dust-enshrouded star formation
episodes (probed by higher redshift SMGs), or if a significant
fraction were formed in less vigorous star-forming systems traced by
higher redshift LBGs.  By determining the number density of massive
galaxies in the LBG phase at each redshift interval, we hope to
estimate the fraction of DRGs whose progenitors formed a significant
component of their mass  in relatively unobscured systems with
moderate star formation rates (SFR) typical  of the LBG population
(e.g., \citealt{Shapley01,Shapley05}).

A further question of significance in characterizing the mode of early
galaxy assembly is understanding the duration of star formation
episodes  in high redshift galaxies.  In the context of models in
which  gas accretion dominates galaxy growth at high redshift (e.g.,
\citealt{Birnboim03,Keres05,Finlator07}) leading to rapid and steady
star  formation,  one might imagine that the LBGs seen at $z\simeq 4$
have evolved relatively smoothly since their formation at an earlier
epoch, creating stars at a near constant rate.  If their assembly
timescales are long enough, we would thus expect similarly luminous
$z\simeq 5$ and 6 LBGs to be less evolved versions than their $z\simeq
4$ descendants with lower stellar masses and younger ages.
Alternatively, it is conceivable that this era witnesses a rapid
increase in the intensity of star formation in individual  galaxies,
as has been predicted for the early stages of galaxy growth
\citep{Keres05,Finlator07}.  In this  case, galaxies would grow along
a locus of  points (or ``main sequence'') in the $\rm{M_\star}$-SFR
plane.  Thus, if viewed at fixed UV luminosity, the observed  stellar
populations would not vary significantly from one redshift to another.
Finally, in contrast to these  ``sustained'' star formation histories,
we  may expect star formation episodes to occur on much shorter
timescales.  If the past duration of star formation is sufficiently
short ($\lsim 300$ Myr), then each dropout sample would be dominated
by newly emerged systems, and  we would not necessarily expect to see
significant growth in the $\rm{M_\star}$-SFR plane over the redshift range
studied.

Earlier work in this direction has been promising but the datasets
have  been limited. \cite{Drory05} traced the mass assembly of
galaxies to high redshift using K-band selected samples. However since
their study did not include IRAC data, reliable stellar masses could
only be derived to $z\simeq 4.5$.  \cite{Verma07} used HST and IRAC
data to compare the stellar mass and age distribution of a small,
robust sample of $z\simeq 5$ LBGs to those derived by \cite{Shapley01}
at $z\simeq 3$.  Most recently, \cite{McLure08} computed the stellar
mass function of  LBGs at $z\simeq 5$ and 6 by scaling the UV
luminosity function by the average mass/light ratio of the LBG
population at high redshift.  While these efforts have improved our
understanding of  the stellar content of high-redshift galaxies,  none
has yet systematically traced the evolving stellar populations of
galaxies over the full redshift range, using large and uniformly
selected samples with no scaling assumptions.  

A plan of the paper follows.  In \S2, we describe the GOODS data used
in our  analysis.  In \S3, we discuss the color selection used to
identify  dropouts, and criteria used to remove contaminants from our
sample.  We close  the section by assessing the evolving surface
densities of the dropouts.  In \S4, we consider the mid-infrared
properties of the dropouts and in \S5, we discuss the population
synthesis models used to infer stellar masses and ages and comment on
the uncertainties in the derived properties.  In \S6, we examine the
redshift evolution in the stellar masses and ages of dropouts  at a
fixed UV luminosity.  Using this information, we discuss the
implications for the star formation histories of galaxies at high
redshift.  In \S7, we study the evolving stellar mass functions  of
LBGs over $4\lsim z\lsim 6$ and estimate the fraction of LBGs that
evolve into quiescent massive galaxies at $z\simeq 2-3$.   Finally in
\S8, we compute the stellar mass densities of the B, V, and  $i'$-drop
samples.

Throughout the paper, we adopt a $\Lambda$-dominated, flat universe
with $\Omega_{\Lambda}=0.7$, $\Omega_{M}=0.3$ and
$\rm{H_{0}}=70\,\rm{h_{70}} {\rm km\,s}^{-1}\,{\rm Mpc}^{-1}$. All
magnitudes in this paper are quoted in the AB system \citep{Oke83}.

\section{Data}

\subsection{The GOODS Fields}

We focus our analysis on the data from the Great Observatories Origins
Deep Survey (GOODS).  Detailed descriptions of the datasets are
available in the literature \citep{Giavalisco04a}, so we only provide
a brief summary here.  The GOODS-S and GOODS-N survey areas each cover
roughly 160 arcmin$^2$ and are  centered on the {\em Chandra} Deep
Field South (CDF-S; \citealt{Giacconi02}) and the {\em Hubble} Deep
Field North (HDF-N; \citealt{Williams96}).   Extensive multiwavelength
observations have been conducted in each of these fields.  In this
paper, we utilize optical imaging from the Advanced Camera for Surveys
(ACS) onboard the Hubble Space Telescope (HST).  Observations with ACS
were conducted in F435W, F606W, F775W, and F850LP (hereafter
B$_{435}$, V$_{606}$, $i'_{775}$, $z'_{850}$) toward GOODS-S and
GOODS-N \citep{Giavalisco04a}.  The average 5$\sigma$ limiting
magnitudes in the v1 GOODS ACS data (correcting for the  flux that
lies outside the 0\farcs5 diameter photometric aperture) are
B$_{435}$=27.8, V$_{606}$=28.0, $i'_{775}$=27.2, and $z'_{850}$=27.0.
We also make use of U-band observations of GOODS-N taken with the Kitt
Peak National Observatory 4-m telescope with the MOSAIC prime focus
camera \citep{Capak04} and of GOODS-S taken with the Wide-Field Imager
mounted on the 2.2m MPG/ESO telescope \citep{Arnouts01}. 

In the near-infrared, we utilize publicly available deep J and K-band
observations of GOODS-S (PI: C. Cesarsky) using the ISAAC camera on
the  Very Large Telescope (VLT).  The sensitivities vary across the
field  depending on the effective integration time and seeing FWHM.
Average 5$\sigma$ magnitude limits (corrected for the amount of flux
that falls outside of the 1\farcs0 diameter aperture) are
$J\simeq 25.1$ and K$_s\simeq24.6$.   Toward GOODS-N, we make use of a
new Subaru K-band mosaic of the field \citep{Bundy09} using the MOIRCS
camera \citep{Ichikawa06}.  The specifics of the  observations are
described in detail in \cite{Bundy09}.  The 5$\sigma$ limiting
magnitude varies between $K_s=23.9$ and 24.1 (after aperture
correction) across the mosaic.  

Deep {\it Spitzer} imaging is available toward both GOODS fields with
the Infrared Array Camera (IRAC) as part of the ``Super Deep'' Legacy
program (Dickinson et al. {\it in prep}, Chary et al. {\it in prep}).
Details of the observations have been described in detail elsewhere
\citep{Eyles05,HYan05,Stark07a} so we do not discuss them  further
here.  The 5$\sigma$ limiting magnitudes of the IRAC imaging are
$\simeq 25.9$ at 3.6$\mu$m and $\simeq 25.5$ at 4.5$\mu$m using
2\farcs4 diameter apertures and applying an aperture correction  (see
\S2.3). 

In a later section (\S3.3), we will revisit the sensitivity limits of
the GOODS datasets, utilizing completeness simulations to  determine
magnitude limits for our survey.

\subsection{Optical Photometry}

ACS photometry was obtained from the GOODS team r1.1 catalog~
\footnote{available from {\tt http://archive.stsci.edu/prepds/goods}}
which ws generated from the v1 GOODS ACS reduction.  The photometric
zeropoints  adopted in the catalog  are 25.653, 26.493,  25.641, and
24.843 for the $B_{435}$-band, $v_{606}$-band, $i'_{775}$ band, and
$z'_{850}$-band, respectively.  We have corrected for the small amount
of foreground Galactic extinction using the {\it COBE}/DIRBE \& {\it
  IRAS}/ISSA dust maps of \cite{sfd98}; for the GOODS-S field, the
color excess is given by E($B-V$) = 0.008 mag; in GOODS-N, the color
excess is given by E($B-V$) = 0.012  mag. Colors are computed using
magnitudes measured 0\farcs50-diameter apertures.  As detailed further
in \S3.1, total magnitudes are computed  using a combination of the
aperture colors and MAG\_AUTO SExtractor parameter.

\subsection{Near- and mid-infared Photometry}

Near-infrared fluxes were computed on each of the sources in the GOODS
r1.1 catalog.  We utilized $1''$-diameter apertures centered on the
source positions in the {\it ACS} images.  The seeing varied across
the GOODS-S and GOODS-N fields as different tiles were taken over many
nights, so we determined separate aperture corrections from unresolved
sources for each tile. For the $J$ and $K_s$-band ISAAC images the
seeing is typically good (${\rm FWHM}=0\farcs4-0\farcs5$), and the
aperture corrections are $\approx 0.3-0.5$\,mag, determined from
bright but unsaturated isolated stars measured in $6''$-diameter
apertures.  Likewise, the MOIRCS GOODS-N $K_s$-band seeing was fairly
constant at {\rm FWHM=0\farcs5; photometry was again computed in
  $1''$-diameter apertures and the aperture corrections are typically
  0.3-0.7\,mag.  

For the GOODS IRAC images, magnitudes  are measured in relatively
small apertures ($\approx 1.5\times {\rm FWHM}$ which corresponds to a
diameter of 2\farcs4) to maximize the signal-to-noise ratio ($S/N$).
We applied aperture corrections to compensate for the flux falling
outside the aperture: these were $\approx 0.7$\,mag for the IRAC 3.6
and 4.5 $\mu$m data, as determined from bright but unsaturated point
sources in the images using large apertures.  The PSF of IRAC leads to
frequent blending with nearby sources, so care must be taken to ensure
that the photometry is robust.   We discuss our strategy for dealing
with this in \S4.  

\section{Selection of High-Redshift Galaxies}

The goal of this section is to compile a robust sample of dropouts  at
$z\simeq 4$, 5, and 6.  In \S3.1, we identify B, V, and $i'$-dropouts
using standard selection criteria (e.g.,
\citealt{Giavalisco04b,Beckwith06, Bouwens07}).  We excise
low-redshift and stellar contaminants from these samples using a
combination of photometric,  spectroscopic,  and morphological
techniques (\S3.2).  In \S3.3, we compute the completeness limits of
our data and determine the surface density of our samples. 

\subsection{Dropout Selection}

Galaxies at $z\simeq 4$, 5, and 6 are selected via the presence of the
Lyman-break  as it is  redshifted through the B$_{435}$, V$_{606}$,
and $i'_{775}$ bandpasses, respectively.  Selection of Lyman break
galaxies  at these redshifts has now become routine
\citep{Stanway03,Giavalisco04b,Bunker04,Beckwith06,Bouwens07}.  In
order to ensure a consistent comparison of our samples to these
previous samples, we adopt color criteria which are identical to those
used in \cite{Beckwith06} and very similar to those used by
\cite{Bouwens07}.  These criteria have  been developed to select
galaxies in the chosen redshift interval while  minimizing
contamination from red galaxies likely to be at low redshift.  We
detail our color-selection and S/N limit criteria for selecting
dropouts  below.  

As mentioned in the previous section, galaxies are selected from the
GOODS version r1.1 ACS multi-band source catalogs.  In these catalogs,
source detection has been performed using the $z'_{850}$-band images.

Conditions for selection as B$_{435}$-drop :
\begin{eqnarray}
	B_{435} -Ð V_{606} & > & (1.1 + V_{606} -Ðz'_{850})
        \\  B_{435}Ð- V_{606} & >  &1.1 \\  V_{606} Ð- z'_{850}  & < &
        1.6 \\  S/N(V_{606}) &>& 5 \\   S/N(i'_{775}) &>& 3 
\end{eqnarray}

In addition we also examine each B-drop for a detection in the U-band
images  described in the previous section.  B-band dropouts at $z\gsim
3.5$ should not  show strong detections in the U-band as it lies
shortward of the dropout  filter.  We look at each source with a
U-band  detection (2$\sigma$ or more) to make sure that the flux is
not spurious  or associated with another object.  In total, we excise
four sources based on the presence of U-band flux.  

Conditions for selection as V$_{606}$-drop:
\begin{eqnarray}
	V_{606} Ð- i'_{775}  &>& 1.47 + 0.89(i'_{775} Ð- z'_{850})
        \nonumber \\ &   &  {\rm or}\ \ 2  \\  V_{606} Ð- i'_{775} &>&
        1.2 \\ i'_{775} Ð- z'_{850} &<& 1.3 \\ S/N(z'_{850})      &>& 5
        \\ S/N(B_{435})      &<& 2 \hspace{1cm}  {\rm or}\ \ \nonumber
        \\ B_{435}-i_{775} &>& V_{606} - i'_{775} + 1
\end{eqnarray}

Conditions for selection as $i'_{775}$-drop:    
\begin{eqnarray}
	i'_{775} Ð- z'_{850} &>& 1.3 \\  S/N(z'_{850}) &>& 5
        \\  S/N(V_{606}) &< & 2 \hspace{1cm}  {\rm or}\ \ \nonumber
        \\ V_{606}-z'_{850} &>& 2.8
\end{eqnarray}

In order to utilize these dropout criteria, we must compute accurate
colors and total magnitudes for each source in the GOODS catalogs.
The  colors are computed using the aperture magnitudes discussed in
\S2.2.   We adopt the $i'_{775}$-band MAG\_AUTO magnitude as the total
$i'_{775}$-band  magnitude for the B$_{435}$-drops. Total magnitudes
are subsequently  computed in the B$_{435}$, V$_{606}$, and
$z'_{850}$-bands using the  B$_{435}-i'_{775}$, V$_{606}-i'_{775}$,
$i'_{775}-z'_{850}$ colors  measured using aperture magnitudes
(0\farcs50-diameter).  Similarly, total magnitudes  are computed for
the V$_{606}$ and $i'_{775}$-drops taking the $z_{850}$-band MAG\_AUTO
magnitude and B$_{435}-z'_{850}$, V$_{606}-z'_{850}$,
$i'_{775}-z'_{850}$ colors.  The S/N limits stated above are
determined  from the measured flux and error in the total magnitudes
and vary depending  on the noise background at the location of the
source.  On average,  these limits are well represented by the
magnitudes presented in \S2.2.

\begin{figure}
\figurenum{1}  \epsscale{1.0} \plotone{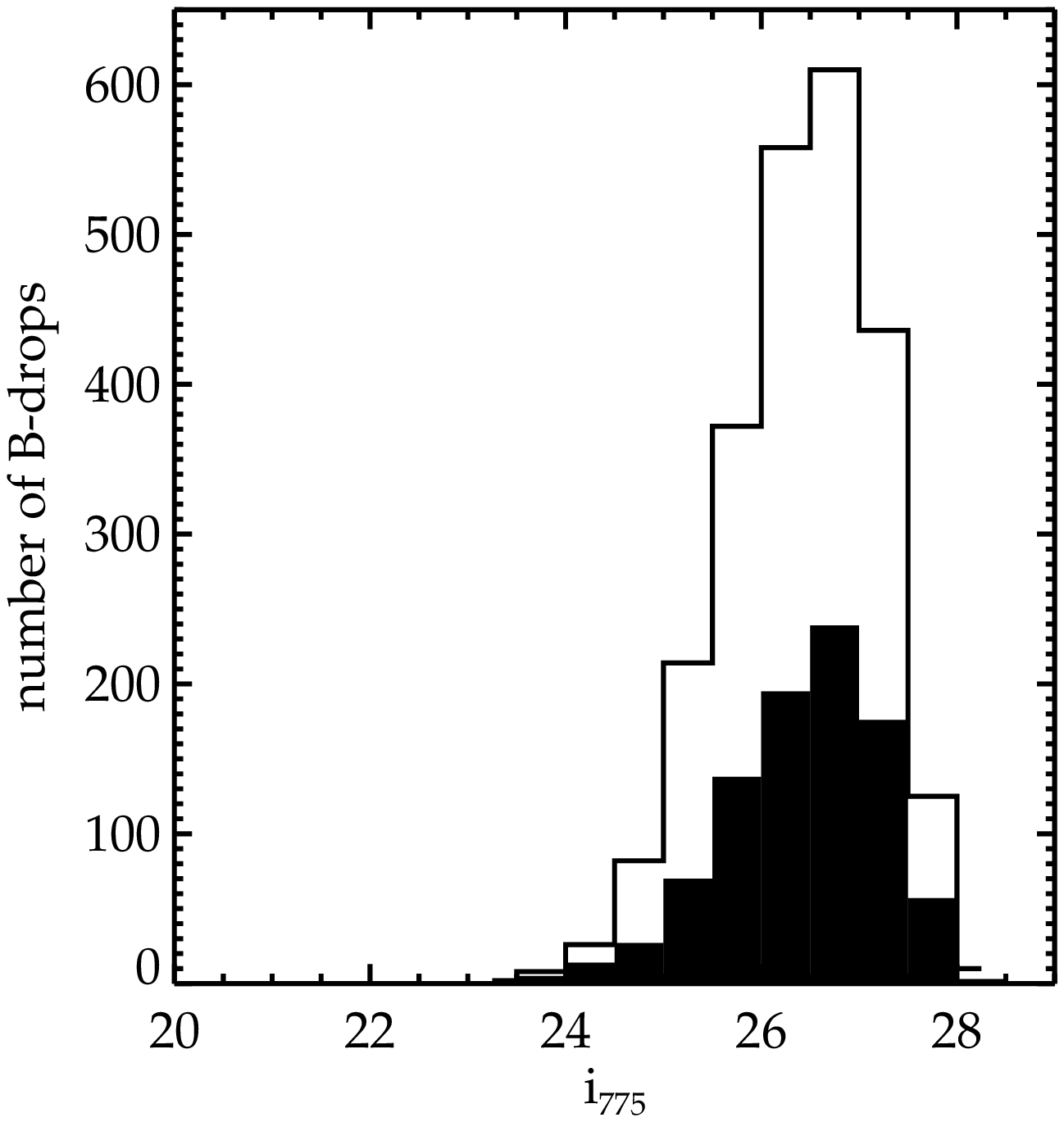} \plotone{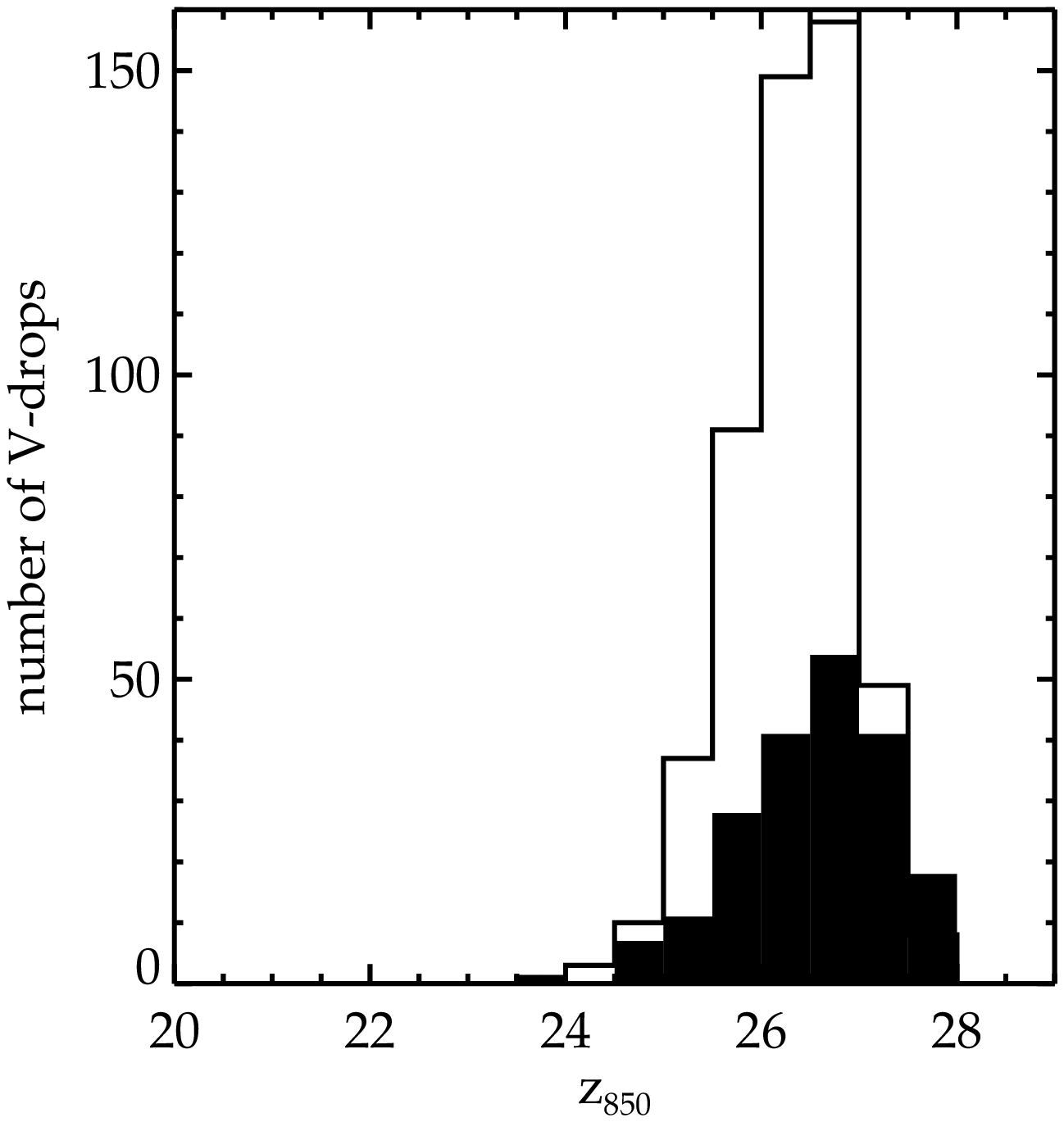}
\plotone{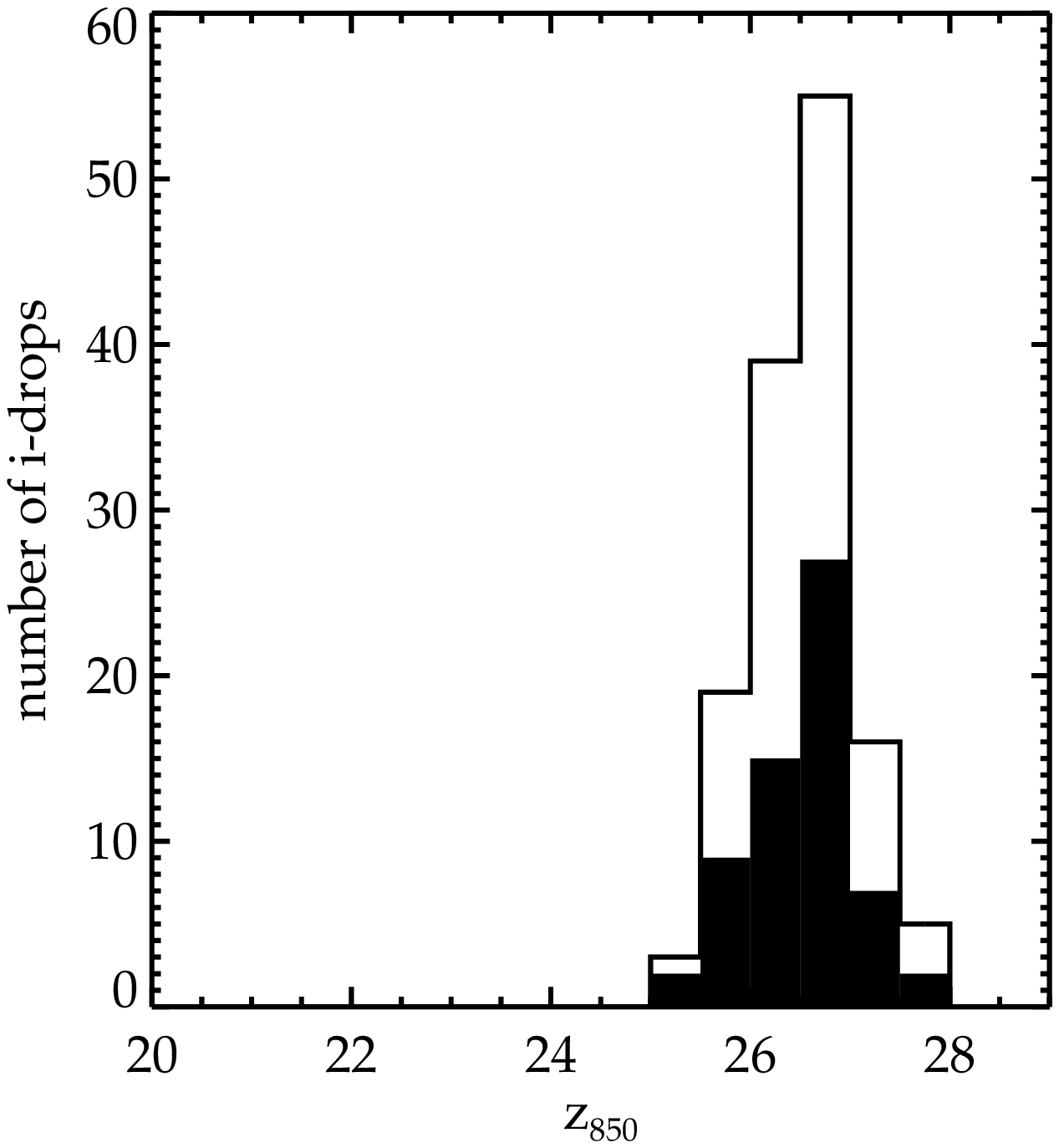}
\caption{Histogram of $i'_{775}$ and $z'_{850}$ magnitudes for 2819 B-drops, 615
  V-drops,  and 166 $i'$-drops in GOODS-S  and GOODS-N. In each panel,
  the open histogram corresponds to the total  sample of dropouts
  while the filled histogram corresponds to only  those sources that
  are isolated in Spitzer (see \S4.1).}
\end{figure}

We examine each source in the dropout catalogs, removing diffraction
spikes from bright stars, image artifacts, and spurious features
(primarily along the edges of the GOODS mosaic), leaving a sample of
2819 B-drops,  615 V-drops, and 166 $i'$-drops.  A histogram of the
measured optical magnitudes for the three dropout samples is presented
in  Figure 1.  In the next section, we seek to further refine these
samples, identifying and removing stellar and low-redshift
interlopers.

\subsection{Removal of Stellar and Low-z Contaminants}

The Lyman-break selection is well known to include a variety of
interlopers in addition to the desired high-redshift galaxies.
Standard contaminants include cool stars and dusty or old  galaxies at
lower-redshifts (e.g., \citealt{Stanway04,Beckwith06,Bouwens06a}). We
first seek to morphologically identify and excise stellar interlopers
from the dropout catalogs.  To assess the fraction of dropouts at low
redshift, we first mine the existing spectroscopic surveys for known
redshifts of our sample.  We then proceed to compute photometric
redshifts for all the remaining dropouts without spectroscopic
redshifts and compute the uncertainty in the photometric redshifts.  

We identify stellar contaminants by their morphology, removing all
bright ($z'_{850}< 26.0$) sources with SExtractor stellarity parameter
greater than 0.80.  We found that this value (very similar to those
used in Bouwens et al.) was optimal in distinguishing between
spectroscopically  confirmed galaxies and stars.  Applying this
criteria removes 51 B-drops (20 in GOODS-S, 31 in GOODS-N), 23 V-drops
(12 in GOODS-S, 11 in GOODS-N), and 8 $i'$-drops (3 in GOODS-S, 5 in
GOODS-N) from our sample.  Faintward of this limit, the S/N of the
GOODS data is too low to reliably identify stars by their stellarity
index.  Following the approach taken in \cite{Bouwens06a}, we estimate
the number of faint stellar contaminants that remain in the GOODS
dropout catalog by examining dropout samples in the Hubble Ultra Deep
Field (HUDF); the exquisite data quality of the HUDF enables point
sources to be identified with sufficient S/N for sources that would be
located at the sensitivity limits of the GOODS data.  Using the same
dropout criteria as in GOODS, we select B, V, and $i'$-drops in the
HUDF using publicly-available $z_{850}$-band source catalogs
\citep{Coe06}.  We find that the fraction of dropouts with stellarity
indices above 0.8 over the magnitude interval $26<z_{850}<28$ is very
low  for the B-drops (0/171), V-drops (0/35), and $i'$-drops (0/21).
These results support the findings of previous studies
(e.g. \citealt{Bunker04, Pirzkal05, Bouwens06a, Bouwens07}) that
stellar contaminants are most prevalent at relatively bright
magnitudes (e.g., $z_{850}<26$).

Dropouts in the GOODS fields have been observed spectroscopically with
VLT/FORS2 \citep{Vanzella02,Vanzella05,Vanzella08, Vanzella09},
Gemini/GMOS Nod\&  Shuffle \citep{Stanway04,Stanway07} , and
Keck/DEIMOS \citep{Stanway04}.  The results of the VLT/FORS2 survey
are  made public in a large database of 1165 total spectroscopic
redshifts  in GOODS-S \citep{Vanzella08}.  We search the database
(version 3.0) for each of the dropouts using an 0.5-arcsecond matching
diameter.  Thirty-seven B-drops have spectroscopic redshifts ranging
between $z$=3.19 and 4.73 with a median redshift of 3.7.   We remove
the one B-drop with a low-redshift ($z$=1.54) identification  from our
sample.  Twenty-four V-drops have spectroscopic redshifts at $z>4$
with a median redshift of 4.81.  Finally, 20 $i'$-drops have
spectroscopic redshifts at $z>5$, while one $i'$-drop (which we
subsequently remove) has a low-z identification ($z$=1.32).

Photometric redshifts and their associated probabilities have been
determined for each remaining dropout by fitting the observed SEDs
against template galaxy SEDs of various ages, spectral types,  and
redshifts using the Bayesian Photometric Redshift (BPZ) software
\citep{Benitez00}.    We fit the observed  optical through near-IR SED
against templates from \cite{Coleman80} and the starburst templates
from \cite{Kinney96} and a very young (10 Myr) starburst template
derived from the Charlot \& Bruzual 2007 (hereafter CB07) models
assuming an exponentially decaying star formation history with
$\tau$=100 Myr, no dust,  and a metallicity of 0.2 Z$_\odot$. The code
outputs both a Bayesian and maximum-likelihood photometric redshift
(and associated probability).

\begin{figure}
\figurenum{2} \epsscale{1.2}\plotone{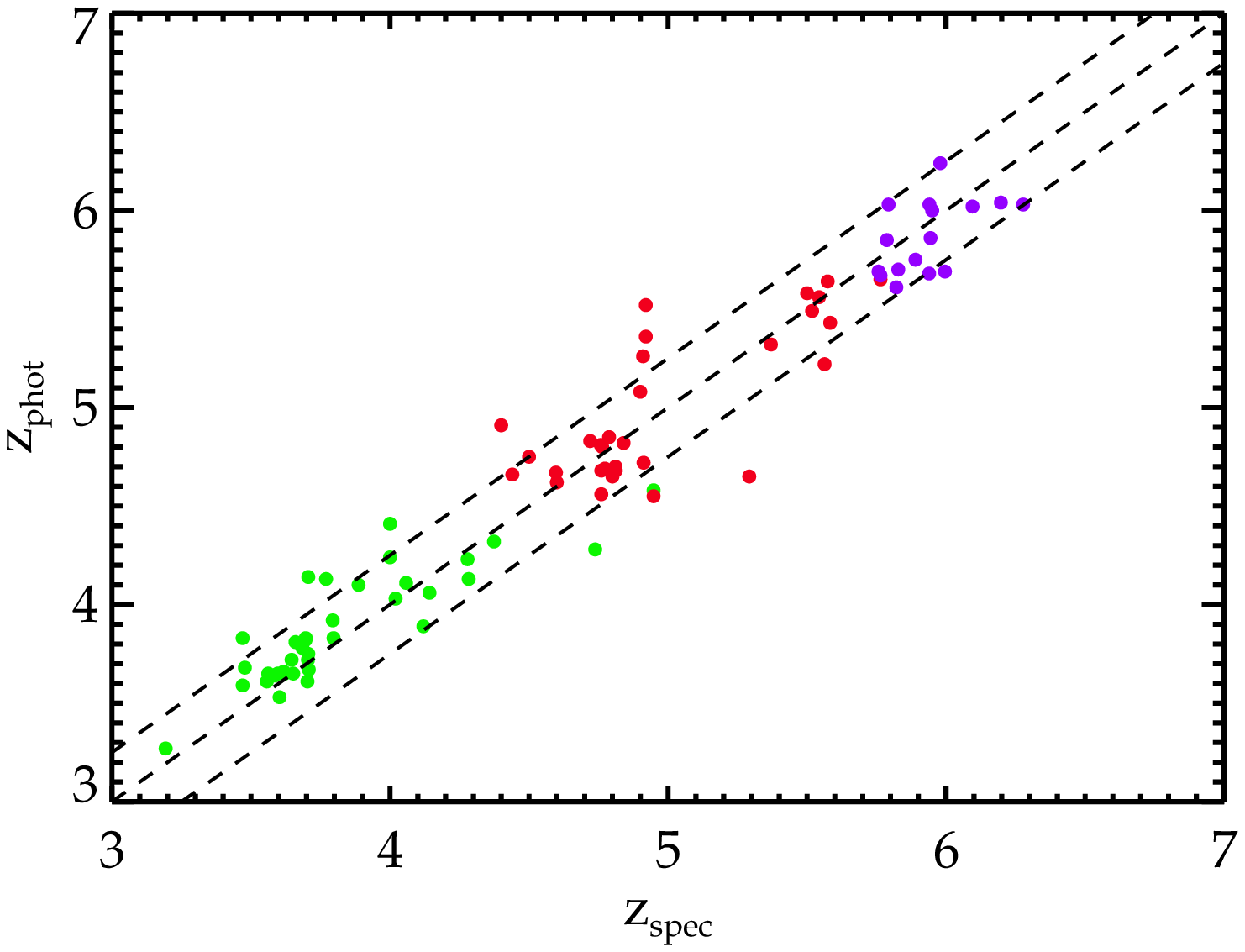} \plotone{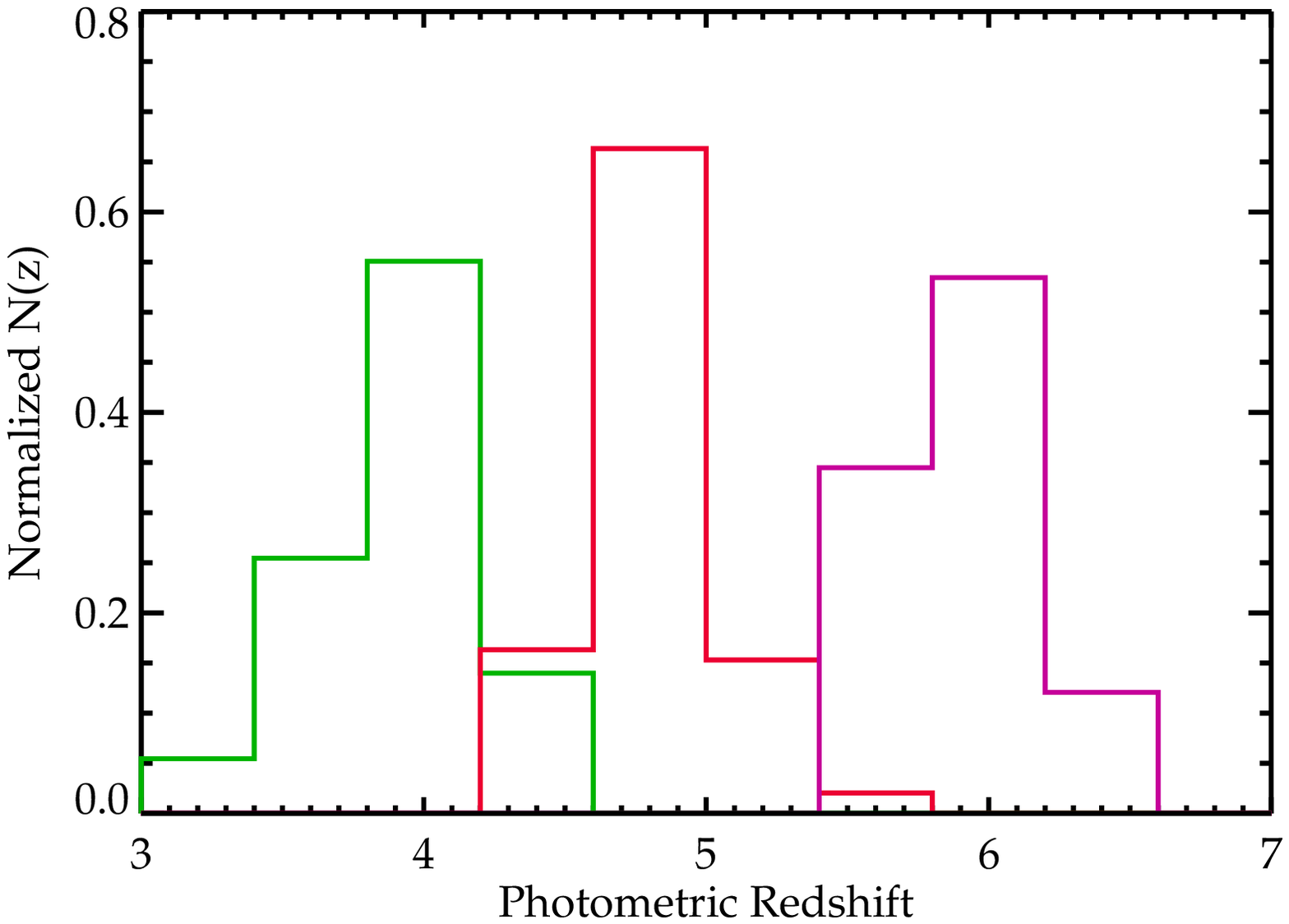}
\caption{{\it Top:} Comparison of photometric and spectroscopic
  redshifts  for dropouts with redshifts measured with FORS2 from
  \citep{Vanzella08}.   The center dashed line denotes the unity
  relationship, while the two additional  dashed lines correspond to
  $\Delta$z=z$_{spec}$-z$_{phot}$=0.25.   {\it Bottom:} Distribution
  of photometric redshifts in B, V, and $i'$-drop  IRAC-clean samples,
  normalized to the total sample size.  The median inferred redshift
  is 3.96, 4.75, and 6.01  for the B, V, and $i'$-drop samples,
  respectively.  }
\end{figure}

In Figure 2a, we present the derived maximum-likelihood photometric
redshifts for the sample of dropouts with known spectroscopic
redshifts  from FORS2.  The root mean square (RMS) difference between
the photometric  and spectroscopic estimates is 0.19, 0.25, and 0.18
for the  B-drops,  V-drops, and $i'$-drops, respectively, excluding
one spectroscopically-confirmed source with an anomalously low
photometric redshift.   We find that the RMS uncertainty in our BPZ
photometric redshifts is similar to that from the  GOODS MUSIC catalog
\citep{Grazian06} for the subset of  dropouts with spectroscopic
redshifts.  

We next attempt to eliminate lower redshift contaminants which made
their way through the initial  color cut, removing those sources with
photometric redshifts that lie below z$_{\rm{phot}}<3$, $4$, $5$ in
the B, V, and $i'$-drops. In order to avoid removing  viable
high-redshift sources, we only remove those sources whose summed
redshift probability distribution suggests that they are very likely
($\gsim 70$\%) to lie at low redshift.  Sources with near equal maxima
at low and high redshift can not be confidently included or excluded.
However these sources are small in number.  The inclusion or exclusion
of this subset with ambiguous photometric redshifts does not affect
the observed optical or infrared  magnitude distribution.  Following
this appraoch, the BPZ photometric redshifts require the removal  of
258, 47, and 14 (10, 8, and 9\%) of the B, V, and $i'$-drop candidates
from the catalogs, leaving a sample of 2443, 506, and 137 B, V, and
$i'$-dropouts  across both GOODS fields.  In general, the excised
sources are either marginally detected in the optical or very red in
their $z'_{850}-K$ colors, as would be expected for low-redshift
sources which satisfy the dropout criteria.  

In Figure 2b, we present the distribution of photometric redshifts
from the remaining high redshift samples in bins of $\Delta$z=0.4.
The  median and standard deviation photometric redshifts of the
resulting B,  V, and $i'$-drop samples are $z_{phot}$=3.96$\pm$0.29,
4.79$\pm$0.25, and  6.01$\pm$0.25 respectively.  These distributions
are in rough agreement  with those computed in \citep{Bouwens07}.

\subsection{Surface Densities and Effective Volume}

The GOODS datasets allow sources to be selected down to magnitudes
fainter than $m_{AB}=27$; however, as is apparent in Figure 1, the
completeness begins to decline brightward of this limit.  Hence,  to
compute the correct density of dropouts, we must account for this
variation in completeness with apparent magnitude.  

To do this, we define a  function $p(m,z)$ that represents the
probability that a galaxy with apparent magnitude m and redshift  $z$
is recovered in our dataset.  We compute this function for each of the
three dropout samples by putting thousands of fake galaxies into the
GOODS images and recreating a photometric catalog for the new image
using the selection parameters  described above. The apparent
magnitudes of the fake galaxies span $i'_{775}=21-28$ (B-drops),
$z'_{850}=22-28$ (V-drops), $z'_{850}=23-28$  ($i'$-drops) and their
redshifts span the interval 3.0$<z<$4.9 for the B-drops, 4.0$\leq
z\leq$5.9 for the V-drops, and 5.0$<z<$7.6 for the $i'$-drops in steps
of $\Delta z$=0.1.  The sizes of the fake galaxies are chosen to be
consistent with distribution of half-light radii derived for $z\simeq
4-6$ galaxies in \cite{Bouwens04c}.  The colors of the fake galaxies
depend on the galaxy redshift and spectral energy distribution (SED).
In order to compute these colors, we use the SED of a CB07 model with
constant star formation history, an age of 100 Myr, and no dust as the
intrinsic rest-frame SED of the fake galaxies. Allowing for a younger
age or a color excess of E(B-V)=0.1 in the fake galaxies' SEDs alters
the completeness by roughly 5\%,  which would not significantly change
any of our results.  The colors are computed at each redshift after
applying the appropriate IGM absorption (e.g. Madau 1995) to the SED.
The probability function, $p(m,z)$ is then given by the fraction of
fake galaxies with apparent magnitude, m, and redshift, $z$, that
satisfy the dropout color selection criteria  defined in \S2.4. 

We next determine the volume observed as a function of apparent
magnitude for each of the  dropout samples following the approach of
\cite{Ste99} using
\begin{equation}
V_{\rm eff}(m)=\int dz\,p(m,z)\,\frac{dV}{dz}
\end{equation}
where $p(m,z)$ is the probability of detecting a galaxy at redshift
$z$ and apparent magnitude $m$, and $dz\,\frac{dV}{dz}$ is the
comoving volume per unit solid angle in a slice $dz$ over the redshift
interval  3.0$<z<$4.9 for the B-drops, 4.0$\leq z\leq$5.9 for the
V-drops, and 5.0$<z<$7.6 for the $i'$-drops. 

\begin{figure}
\figurenum{3} \epsscale{1.0}  \plotone{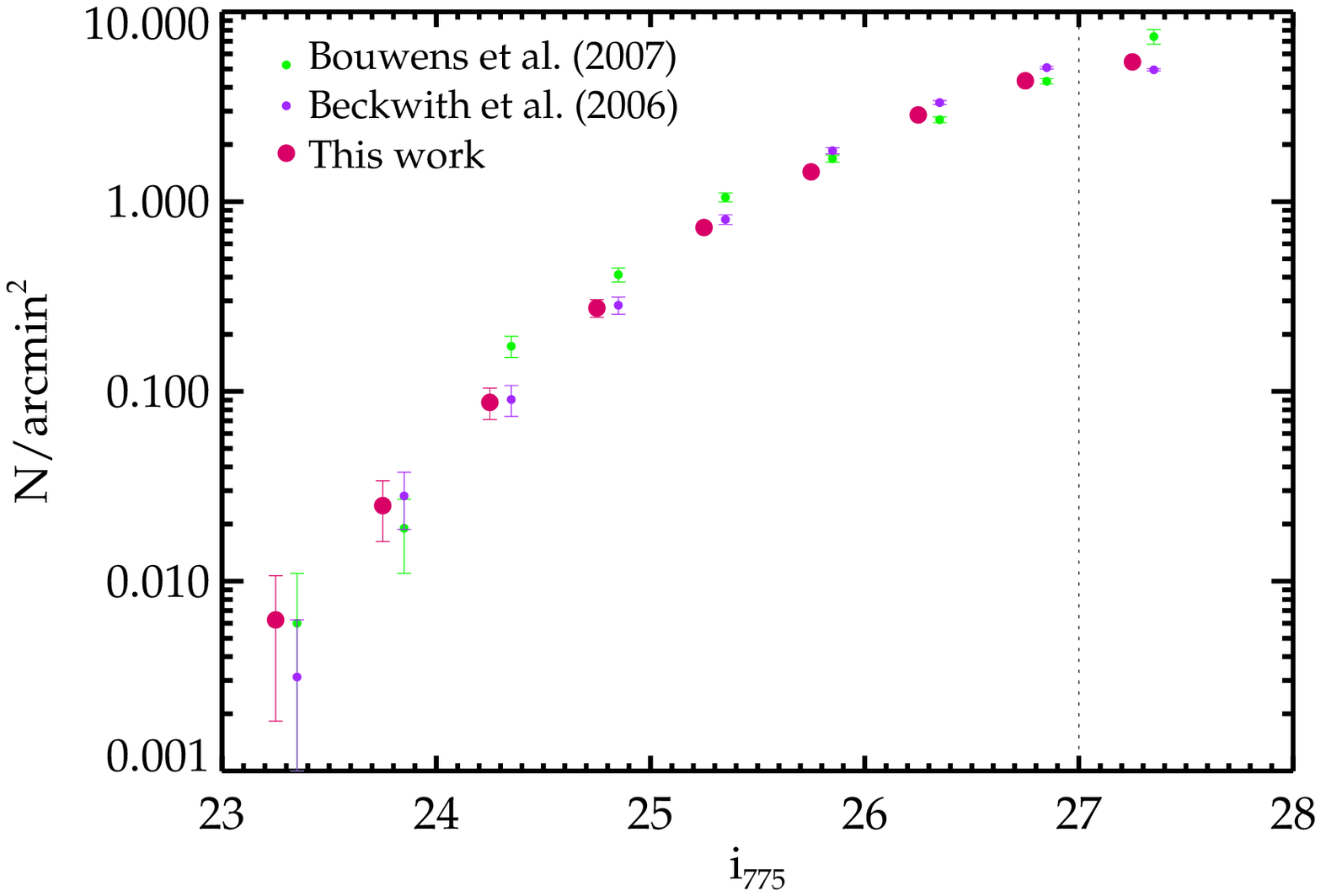} \plotone{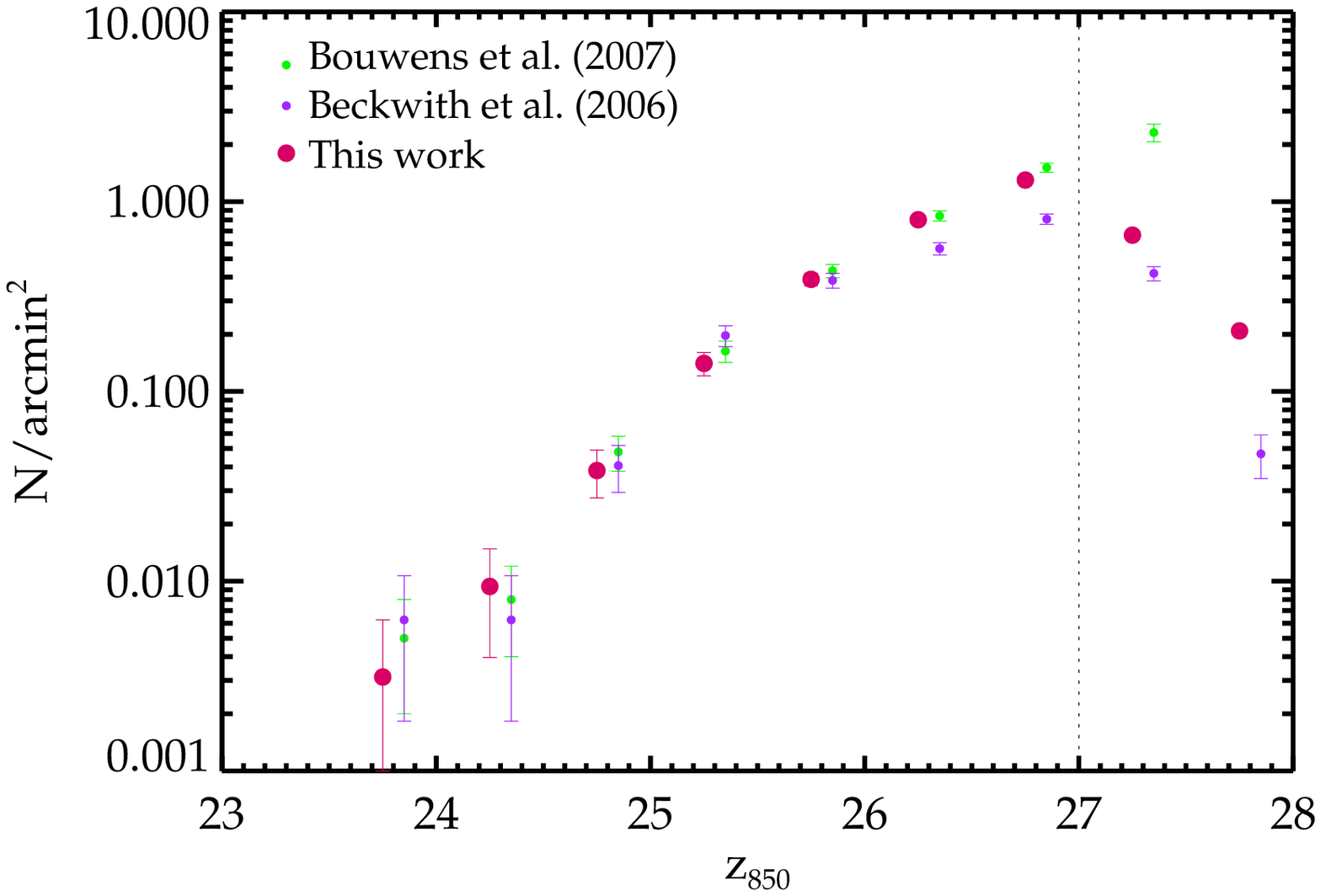}
\plotone{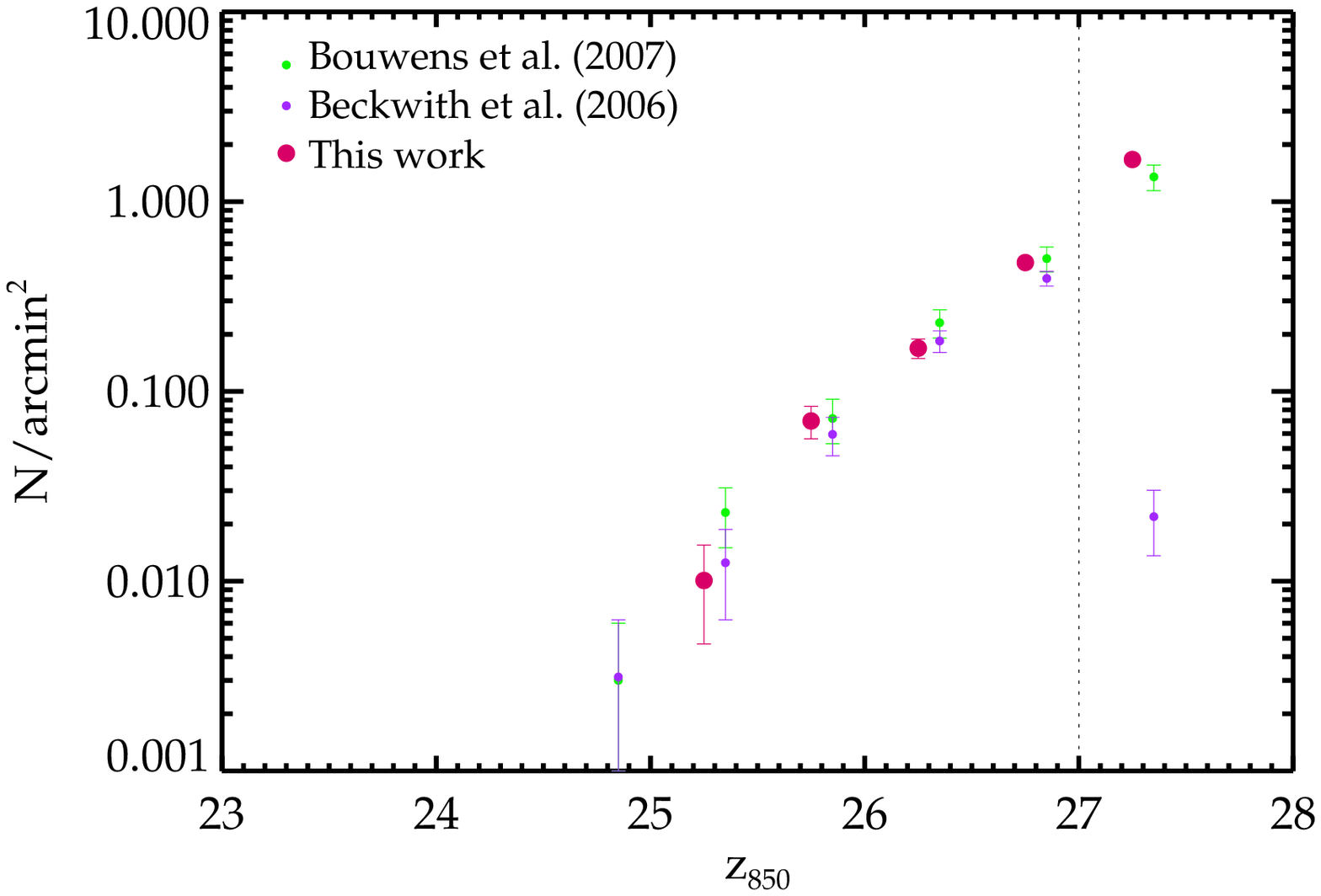}
\caption{Surface densities of dropout samples.  Stars, artifacts, and
  low-$z$ galaxies have been removed from the initial sample as
  described in \S3.2.   We place the remaining sample in magnitude
  bins spanning 0.5 mags and adjust the measured densities for
  incompleteness (\S3.3).  Our samples are sufficiently complete
  brightward of 27th magnitude (vertical dotted lines) to enable
  reliable density measurements.  The resulting surface densities are
  in excellent agreement with previous measurements.  }
\end{figure}

The effective volumes are subsequently computed as a function of
apparent magnitude for the dropout samples following equation 14.  In
general,  for the dropout samples in a single GOODS field, the
effective volumes drop from  4-$5\times \rm{10}^5$ Mpc$^3$ at bright
magnitudes ($m_{\rm{AB}}$=23.5) to $2\times \rm{10}^5$ Mpc$^3$ at fainter
magnitudes ($m_{\rm{AB}}$=27), reflecting $\simeq 50$\% completeness,
on average, at the limit of our samples.  In Figure 3, we show the
derived surface densities of our dropouts, corrected using the
magnitude-dependent completeness $p(m)$ derived in equation above.
The measured values are in excellent agreement (generally to within
1$\sigma$) with previous  measurements \citep{Bouwens07} at AB
magnitudes brighter than 27.  Faintward of this limit, the
completeness in the GOODS data is too low to enable  reliable density
measurements.  

\section{Mid-infrared Properties of Dropouts}

\subsection{Construction of Spitzer-Isolated Subsample}

Mid-infrared photometry is the crucial component for  estimating the
stellar populations of $z\gsim 4$ galaxies. However the low spatial
resolution of {\em Spitzer} results in frequent blending between
nearby sources, making accurate photometry of individual objects
difficult.  To ensure that objects in our sample were not contaminated
by neighboring bright foreground sources, we examined each dropout by
eye and flagged sources that are hopelessly confused with flux from
bright, nearby sources.  

Following \cite{Stark07a}, we only apply population synthesis models
to those objects which are unconfused in the IRAC images, which we
define as having no bright ($\lsim 24$-$25$\,mag), contaminating
sources within a $\simeq 2.5$ arcsecond radius around the position of
interest.  We find that $\simeq 35$\% of the dropouts in our catalogs
are sufficiently isolated  to allow accurate photometry.  In Figure 1,
we show the distribution of $i'_{775}$ and $z'_{850}$ magnitudes  of
the dropouts in our ``Spitzer-isolated'' subsample compared to  the
full dropout samples.  The magnitude distribution of this subsample is
very similar to that of the parent distribution albeit with a slightly
greater proportion of UV bright sources.  We correct for the small
bias this introduces into the stellar mass functions  in \S7.

\begin{figure}
\figurenum{4} \epsscale{1.1} \plotone{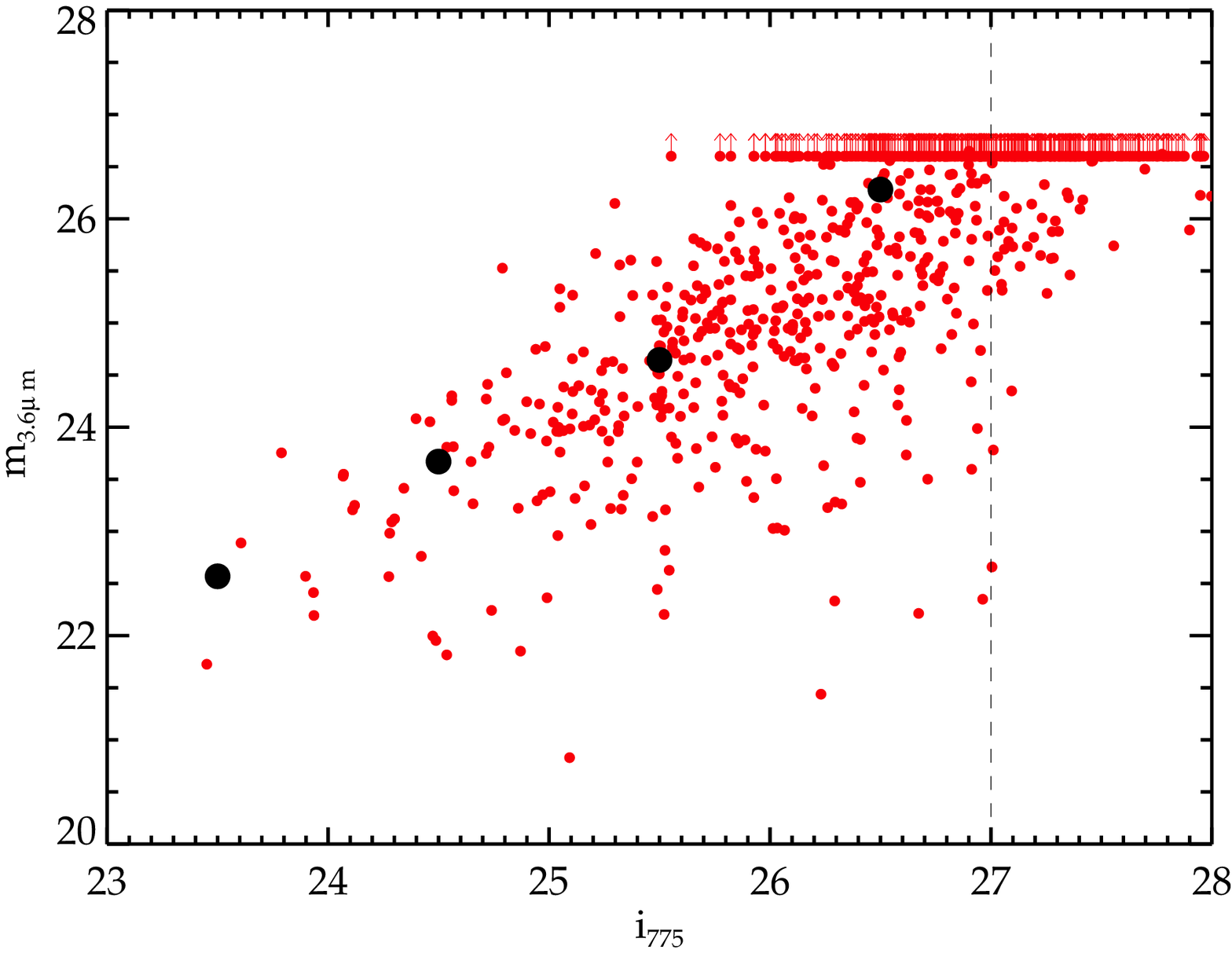} \plotone{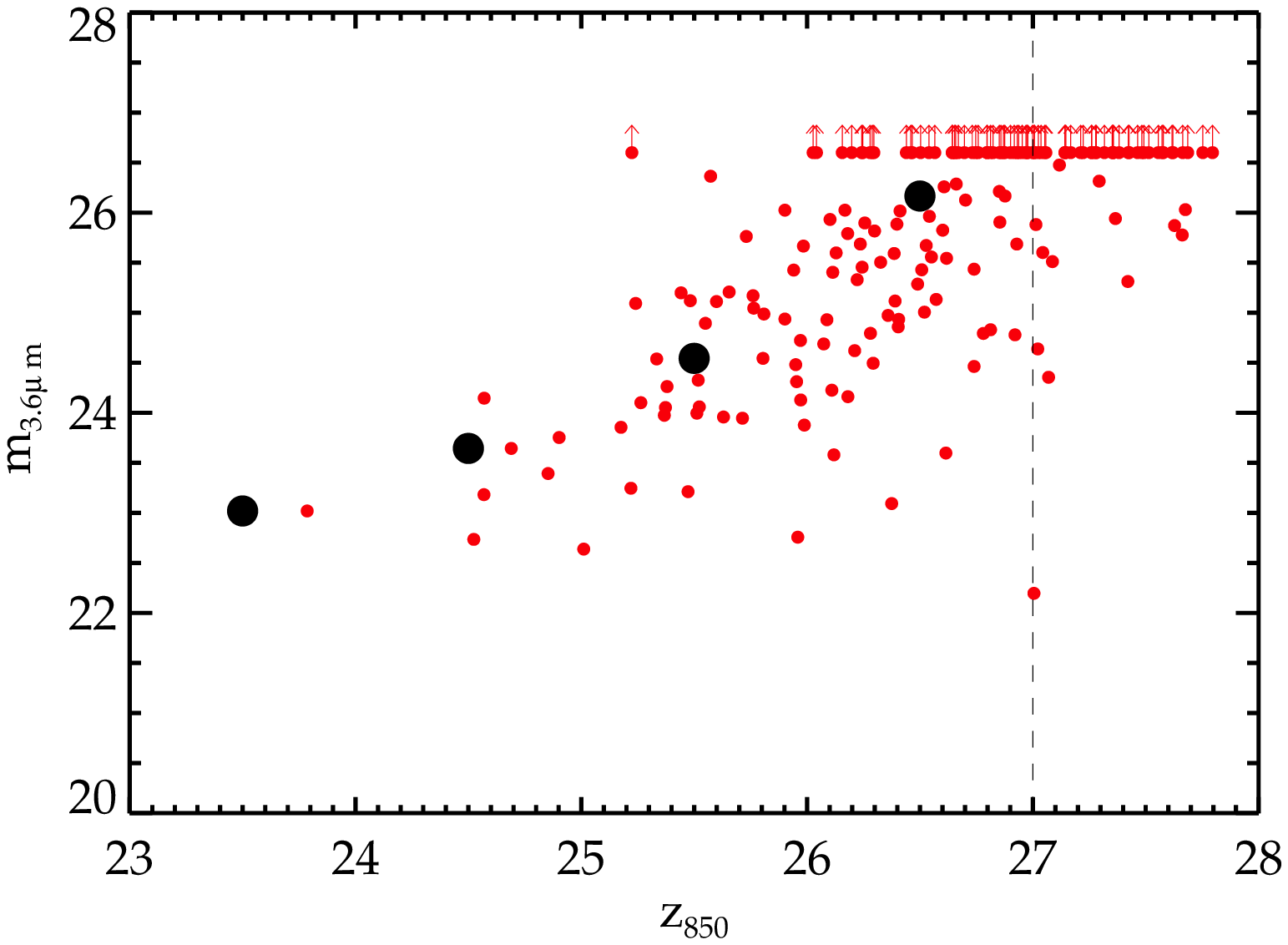}
\plotone{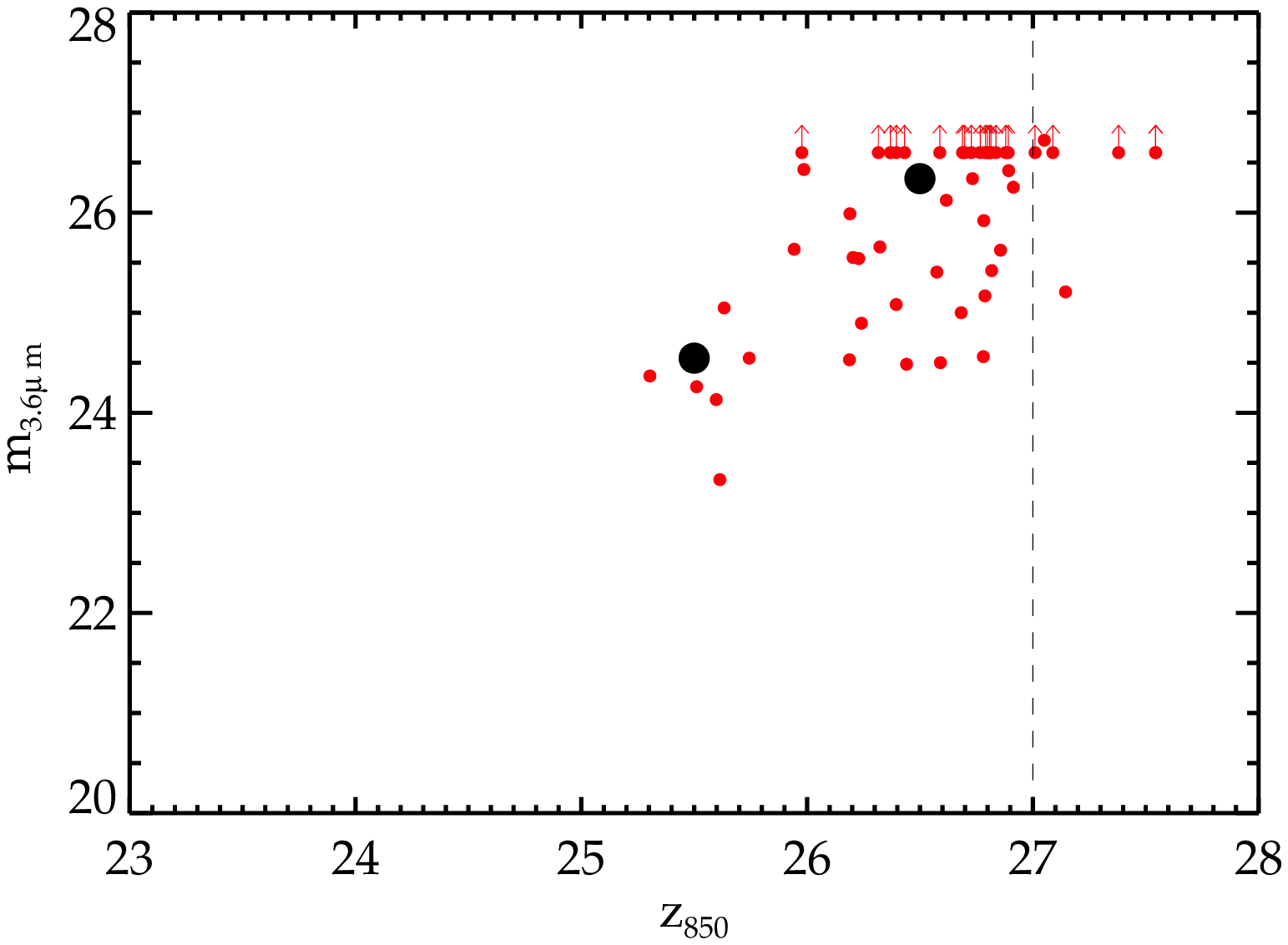}
\caption{Mid-infrared vs. optical flux for the B (top panel), V
  (middle panel), and $i'$-drops (bottom panel).  Individual galaxies
  are shown as small red circles; the large black circles correspond
  to the median 3.6$\mu$m magnitude in bins  of $i_{775}$ or $z_{850}$
  magnitude.  The dashed vertical lines correspond  to our adopted
  completeness limit for each dropout sample. Those objects that are
  brighter in the optical are on average brighter  in the
  mid-infrared.   The relationship between mid-infrared and optical
  flux is very similar for each of the three different redshift ranges
  considered.  }
\end{figure}

In Figure 4, we present the relationship between the IRAC 3.6$\mu$m
magnitude  and $i_{775}$ or $z_{850}$ magnitude for each of our
dropout samples, foreshadowing the relationship between the assembled
stellar mass and the current star formation rate (uncorrected for
extinction).  It is not necessarily obvious that the ongoing star
formation in any galaxy should bear any relation to the past star
formation history, yet immediately apparent is a correlation between
the average optical and mid-infrared flux: sources that are brighter
in the ACS bandpasses are, on average, brighter in IRAC.  Also
noticeable is that the median IRAC flux for a given $i'_{775}$ or
$z'_{850}$ flux does not change significantly  over the redshift range
probed by the three dropout samples.  We discuss the implications of
these trends in more detail in \S6.

\subsection{MIPS Detections}

We have put considerable effort into removing low-redshift
contaminants  from our dataset.  However, as is evident from Figure 4,
the dropout  samples still contain red objects, some of which are
among the  brightest sources detected in IRAC.  Considering the fact
that very massive sources (with bright IRAC fluxes) are likely much
more common at $z\simeq 2$ than at $z\gsim 4$, it is clear that these
sources require more scrutiny before proceeding.

We can gain some insight into the likely redshifts of this population
from 24 $\mu$m imaging with the Multiband Imaging Photometer for
Spitzer (MIPS) camera (Dickinson et al. in prep.; Chary et al. in
prep.).   At $z\simeq 2$, the MIPS imaging passband probes the bright
rest-frame 7.7 $\mu$m feature from polycyclic aromatic hydrocarbons
(PAHs).  As a result, dusty star-forming galaxies at  $z\simeq 2$ are
commonly detected with MIPS (e.g., \citealt{Reddy06}).   Such PAH
features would not be detected in sources over the redshift range our
dropouts sample (4$\lsim z\lsim$6); hence if any of our sources are
detected with MIPS, it very well may indicate that they lie at
$z\simeq 2$.  

In order to determine what fraction of our catalog is detected  at
24$\mu$m, we visually examine the MIPS data of each dropout in the
{\it Spitzer} isolated sample.  While the total number of dropouts
with  MIPS detections is small (12/800 B-drops, 3/186 V-drops, and
none of the $i'$-drops), it is not negligible.  As expected, each of
the sources detected with MIPS is quite red ($z'_{850}-m_{3.6}\gsim
2$) in addition to being bright in the IRAC bandpasses.  These sources
thus make up a significant fraction of the subset of our dropout
samples with bright IRAC fluxes (10/25, 2/4 of those with m$_{3.6}<$23
for the B and V-drops respectively) and are hence sure to strongly
affect  attempts to derive the numder density of massive galaxies.

While we consider these sources as prime low redshift candidates, it
is possible that some of these lie at $z\gsim 4$.  Cross-correlating
the 24$\mu$m-detected subset  with our spectroscopic sample, we find
that  one of the three MIPS-detected V-drops has a spectroscopically
confirmed redshift of $z$=4.76.  Since there are few strong PAH
features that fall into the 24 $\mu$m filter at this redshift, we
propose that the 24 $\mu$m emission most likely comes from a dusty
AGN, a conclusion consistent with the point-like morphology in the
observed optical-frame.  Importantly, this establishes that not all
24$\mu$m-detected dropouts are foreground objects.  The MIPS-detected
subsample thus places an upper limit (1.5\%  for the B-drops and 1.1\%
for the V-drops) on the number of dusty low-$z$ interlopers remaining
in  our samples.  In subsequent sections, we will derive the evolving
stellar populations of our dropout sample both with and without the
24$\mu$m detected sources.

\section{Derivation of Physical Properties}
\label{sec:mass}

We infer stellar masses for the dropout sample by fitting the latest
CB07 stellar population synthesis models to the observed SEDs. These
models include an improved treatment of the thermally pulsing
asymptotic  giant branch (TP-AGB) phase of stellar evolution,
utilizing  the \cite{Marigo07} evolutionary tracks and the
\cite{Westera02} stellar library.    In this section, we first
describe the CB07 models and detail the fitting process that we use to
infer the stellar masses and ages of the galaxies.  We examine the
effects of the TP-AGB stars on the inferred masses and ages of the
dropouts by comparing the CB07 output to that from the
\cite{Bruzual03} population synthesis models (\S5.2).  We then comment
on the uncertainties in the inferred properties of objects  in our
samples and stack the infrared photometry of the faintest sources to
achieve a more reliable estimate of the average stellar populations of
this  subset.

\subsection{CB07 Models}

The CB07 models utilize 220 age steps from $10^5$ to $2\times
10^{10}$\,yr, approximately logarithmically spaced.  For each source,
we do not include age steps in excess of the age of the universe
implied  by its photometric or spectroscopic redshift.  We
also exclude unphysically-low age steps (e.g., below the  dynamical
timescale for LBGs at high-redshift), which we take to be  $\simeq 20$
Myr adopting measurements of the mean half-light radii
\citep{Bouwens04c} and typical velocity dispersions \citep{Erb06} of
LBGs at high-redshift.

The CB07 models only include templates  with the \cite{Chabrier03} and
\cite{Salpeter55} initial mass functions (IMF).  For consistency with
previous efforts, we choose to focus on the Salpeter IMF in our
analysis.  The effects of utilizing a Chabrier IMF have been discussed
in detail elsewhere, including our previous work
\citep{Eyles05,Stark07a}.  

We use the high resolution (FWHM 3\,\AA ) and 1\,\AA\ pixels
evenly-spaced over the wavelength range of 3300\,\AA\ to
9500\,\AA\ (unevenly spaced outside this range).  From the full range
of metallicities offered by the code, we considered both solar and 0.2
Z$_\odot$ models.  From several star formation histories available, we
utilize the constant star formation history and exponentially decaying
($\tau$) SFR models with e-folding decay timescales  $\tau$=100 and
300 Myr.  We also utilize a model with a linearly  increasing star
formation rate, with a slope set to match the observed  brightening in
the UV LF between $z\simeq 6$ and 4 (0.74 mag at 1500~\AA~ over
$\simeq 620$ Myr).  The CB07 spectra are normalized to an SFR of
$1\,\rm{M_{\odot}}\,{\rm yr}^{-1}$ for the continuous star formation
model.  For the exponential decay models, the galaxy mass is
normalized  to M $\longrightarrow$ 1 M$_\odot$ as t $\longrightarrow$
$\infty$.   

We also consider the effects of dust on the integrated SEDs, adopting
the reddening model of \cite{Calzetti97}\footnote{The Calzetti
  reddening is  an empirical law given in terms of color excess E(B-V)
  with the wavelength dependence of the reddening expressed
  as:\newline $\rm
  k(\lambda)=2.656(-2.156+1.509\lambda^{-1}-0.198\lambda^{-2}+0.011\lambda^{-3})+4.88~for~0.12\mu\rm{m}\le\lambda\le
  \rm{0.63} \mu\rm{m}$, and
\newline
k($\rm\lambda)=[(1.86-0.48\lambda^{-1})/\lambda-0.1)]\lambda+1.73~for~0.63
\mu\rm{m}\le\lambda\le \rm{1.0} \mu\rm{m}$ and the flux attenuation is
given by: \newline
F$\rm_{obs}(\lambda)=F_0(\lambda)10^{-0.4E(B-V)}k(\lambda)$.}.  At
each  age step, the model SED was folded in with each of five
different reddening curves corresponding to color excesses of
E(B-V)=0.0, 0.03, 0.1, 0.3, 1.0.  

For each of the galaxies in our sample, the model SEDs were
interpolated onto each of the filters' wavelength scales after
correcting the latter to the rest-frame wavelengths using the redshift
appropriate for each object. Rather than simultaneously fitting  the
mass, age, dust, and redshift, we fixed the redshift at the  best-fit
photometric redshift specified by BPZ.  Where possible, we instead
utilized spectroscopic redshifts.  We discuss the effects of holding
the  redshift fixed later in \S5.3.  To assess the uncertainty in the
inferred parameters (e.g., mass, age, dust) due to redshift error, we
perform the SED fitting at three redshifts: $z=z_{\rm{phot}}-0.25,
z_{\rm{phot}}, z_{\rm{phot}}+$0.25.  

To account for Ly$\alpha$ forest absorption, we decrement the flux
shortward of 1216 \AA~ (rest-frame) by a factor of
$\exp[-(\rm{1}+z)^{3.46}]$ \citep{Madau95}.  We further assume the
flux shortward of 912 \AA~ (rest-frame) is zero due to Lyman-limit
absorption.  

The resulting template SEDs are folded through each of the filter
transmission curves to produce optical, near-IR, and mid-IR magnitudes
for  each model.  Comparing the observed magnitudes  and their
associated errors\footnote{We increase the observed magnitude errors
  to 0.1 when they are less than this value to account for calibration
  uncertainties.} to the derived model magnitudes, we compute a
reduced $\chi^2$ value for each  set of parameters (age, E(B-V),
normalization). For individual sources,  non-detections were set to
the 1$\sigma$ detection limit and the flux error was set at 100\%  (we
also consider the average properties of this faint  sample via a
stacking analysis in \S5.2)

The fitting routine returned the parameters for each model which were
the best-fit to the broadband photometry (i.e., minimized the reduced
$\chi^{2}$).  In addition to selecting those parameters which provide
the best-fit, we also compute the range of ages, normalizations, and
$E(B-V)$ values that result in an SED with reduced $\chi^{2}$ values
within $\Delta \chi_{reduced}^2$=1 of the minimum value; we take these
as the 1-$\sigma$ uncertainties associated  with each parameter.  This
normalization was then converted to the appropriate units and
subsequently used to calculate the corresponding best-fit stellar and
total mass (and the corresponding 1$\sigma$ errors).  The conversion
factor between normalization and mass for a given age and star
formation history was obtained from the '4color' files in the CB07
population synthesis output.

\subsection{The Effect of TP-AGB stars on Inferred Properties at $z\gsim 4$}

The TP-AGB stars contribute significantly to the integrated rest-frame
near-infrared light.  \cite{Maraston05} has shown that if the
contribution from this phase of evolution is  underestimated for
galaxies  at $z\simeq 2$, the stellar masses and ages inferred from
models are typically overestimated.  Note that this is not necessarily
the case at $z\gsim 4$, since at these redshifts the first two IRAC
channels only probe as far redward as the rest-frame optical, where
the contribution  from TP-AGB stars is less dominant than the
rest-frame near-infrared.  In this section, we investigate the  effect
of the TP-AGB stars on the inferred properties of the B, V, and
$i'$-drops by comparing the model output of CB07 and BC03 models.

\begin{figure}
\figurenum{5} \epsscale{1.1} \plotone{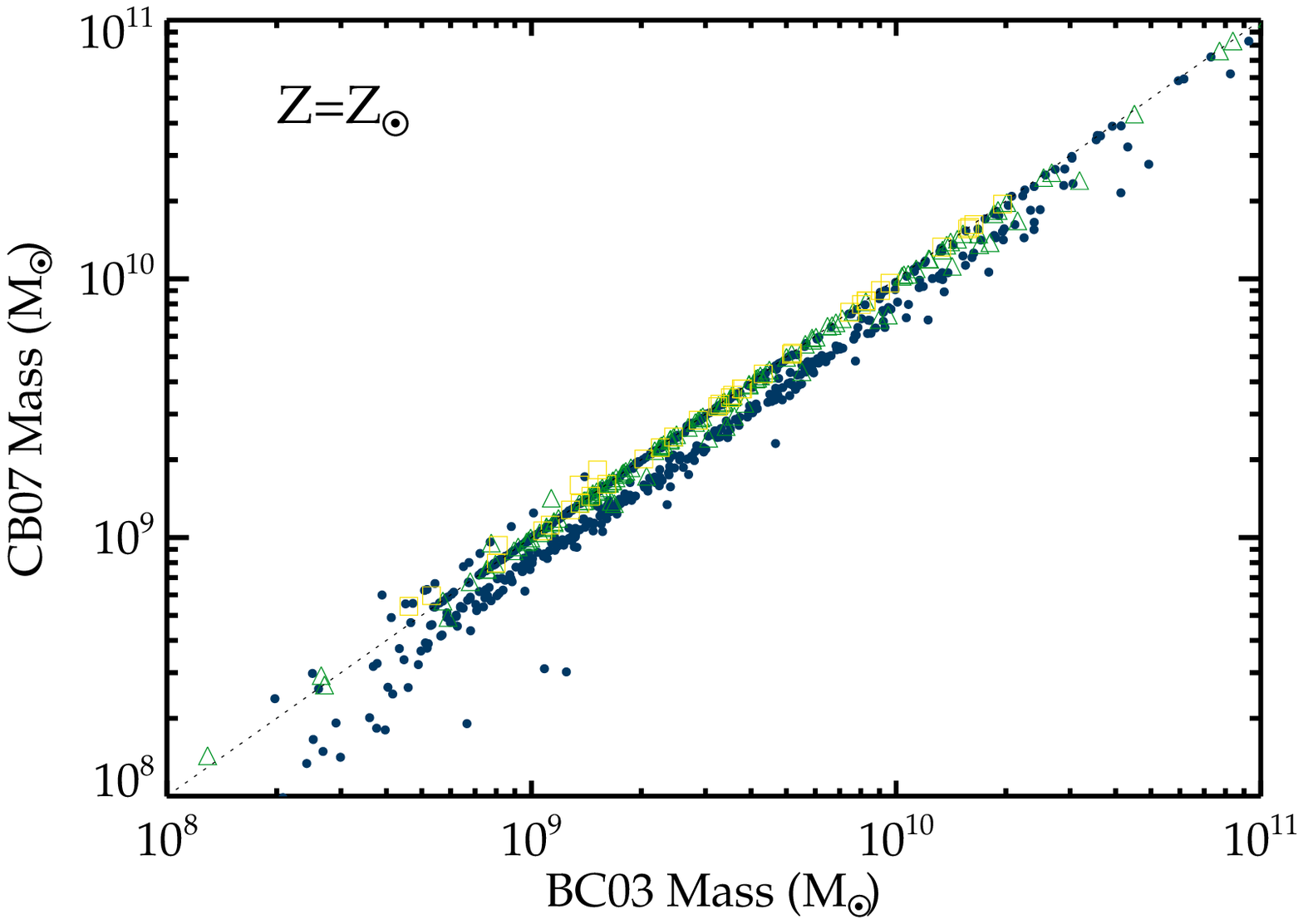} \plotone{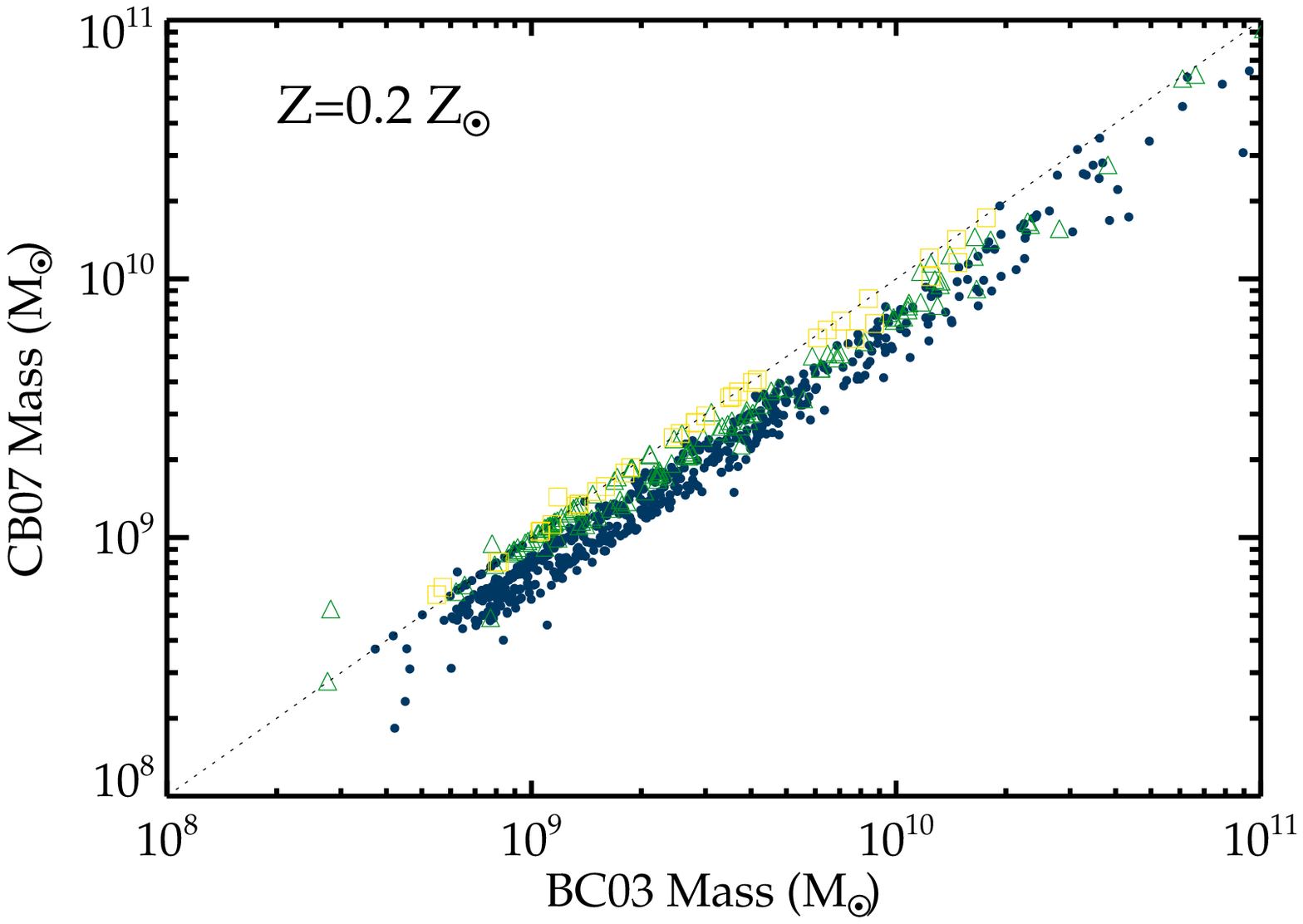}
\caption{Comparison of the model output from \cite{Bruzual03} (BC03)
  and the updated  models presented in \citealt{Bruzual07} (CB07)
  which now include the contribution of TP-AGB stars. The upper panel
  shows B-drops (blue circles), V-drops (green triangles), and
  $i'$-drops (yellow squares) for a solar metallicity $\tau$=100 Myr
  exponential decay model.  The lower panel is identical but with
  sub-solar (Z=0.2 $Z_\odot$) metallicity.  The dotted line in both
  panels  indicates the 1-to-1 relationship between the BC03 and CB07
  mass.  The  discrepancy between the BC03 and  CB07 models increases
  at lower redshifts and lower metallicities.}
\end{figure}

In Figure 5a, the stellar masses of the B, V, and $i'$-drop samples
inferred from BC03 and CB07 are plotted against one other for solar
metallicity.  In computing the masses and ages, we fix the redshift at
the derived photometric redshift and the color excess is fixed
to $E(B-V)$=0. In spite of the addition of TP-AGB stars, the inferred
masses from BC03 and CB07 are in very good agreement. The median
stellar mass of the BC03 $i'$-drops is less than 1\% greater than that
for the CB07 $i'$-drops.  As expected the effect of TP-AGB is more
pronounced at lower redshifts, but nevertheless we still find only a
9\% difference between the median of the BC03 and CB07 B-drop samples.  

At subsolar metallicity, the contribution from TP-AGB stars is
significant ($\gsim $20-30\% of luminosity for 1.3 Gyr instantaneous
burst) in the rest-frame optical \citep{Bruzual07}; hence, we expect
the  CB07 models to return lower masses than the BC03 models for the
0.2 Z$_\odot$  templates.  Indeed, we find that in each dropout sample
the variations between the two models  is more extreme than in the
solar metallicity case (Figure 5b).  In the most discrepant case (the
B-drops), the median mass inferred from  the BC03 models is $\simeq
30$\% greater than that inferred from  CB07.

\subsection{Systematic Uncertainties in Derived Properties}

Before examining the results in detail, we explore the reliability
with which we can determine the stellar populations of $z\gsim 4$
galaxies.  There is ample discussion of the systematic uncertainties
inherent in population synthesis modeling in the literature  (e.g.,
\citealt{Papovich01,Shapley01,HYan05,Eyles05,Shapley05}).  We direct
the reader to these works for detailed discussion and only provide a
brief discussion below on the typical uncertainties in the inferred
masses, ages, and extinction for our dropout samples.

First we must consider whether fixing the redshift at a single value
biases our determination of the mass and its associated  uncertainty.
To test this, we compute the best-fit parameters for the small subset
of sources with spectroscopic redshifts using two different methods:
1) fixing the redshift at the photometric redshift determined by BPZ
and 2) allowing the redshift to float simultaneously with mass, age,
and dust.  Comparing these to the parameters inferred when the
redshift has been fixed at its true spectroscopic value gives some
indication of the error that may be introduced by not allowing the
redshift to float.  We find that the best-fit stellar masses inferred
by method 1 (fixing the redshift at its BPZ photometric redshfit) are
similar to those determined when the redshift is fixed at its
spectroscopic value (the standard deviation in the fractional difference
between the two mass estimates is 17\%).  Varying the redshift
simultaneously  with other parameters  does not result in better
agreement with the masses inferred from the  spectroscopic redshifts
(the standard deviation in the fractional  difference between the two
mass estimates is 19\%).   Moreover, we find that the uncertainty in
the stellar mass that is found by perturbing the redshift by
$\Delta$z=$\pm$0.25 from its best-fit BPZ photometric redshift and
allowing  the dust and age to vary (42\% error)  is very similar to
that determined by allowing the redshift to float simultaneously with
dust and age (48\% error).   These  results suggest that our method of
using fixed redshifts in our SED  fitting analysis will not strongly
bias the mass determinations or  decrease the inferred mass error. 

We now move on to quantify the uncertainty in the derived parameters
for sources in our dropout samples.  The SED fitting procedure is
clearly most  effective at fitting sources that are  bright enough to
be detected with sufficiently high S/N ($\simeq 10$) in the rest-UV
and rest-optical.  For the  bright $z'_{850}=24.3$ dropout presented
in Figure 6 (typical of bright B-drops  in our sample), acceptable
fits (within $\Delta\chi_{reduced}^2$=1  of the best-fit) are found
for models with ages between 180 and 321 Myr, color excesses 
between E(B-V)=0.0 and 0.1, and stellar masses between 2.0 and
2.8$\times$10$^{10}$  M$_\odot$. If we allow the redshift to vary by
$\pm 0.25$ and consider the range of models that produce a reduced
$\chi^2$ within 1 of minimum reduced $\chi^2$ at the central BPZ
photometric redshfit, the inferred stellar masses vary by less than
5\% and  ages change by less than $\simeq 30$\%.  Variations in the
form of the star formation history do change the absolute values
quoted above (see  Shapley et al. 2005 for detailed discussion), with
best fit stellar masses increasing by 35\% and ages reaching 1.4 Gyr
for a  constant star formation model.  Nevertheless the {\it relative}
stellar masses and ages within our samples should be reasonably well
determined for equally bright dropouts, unless the star formation
history varies considerably within the sample. 

\begin{figure}
\figurenum{6} \epsscale{1.0}  \plotone{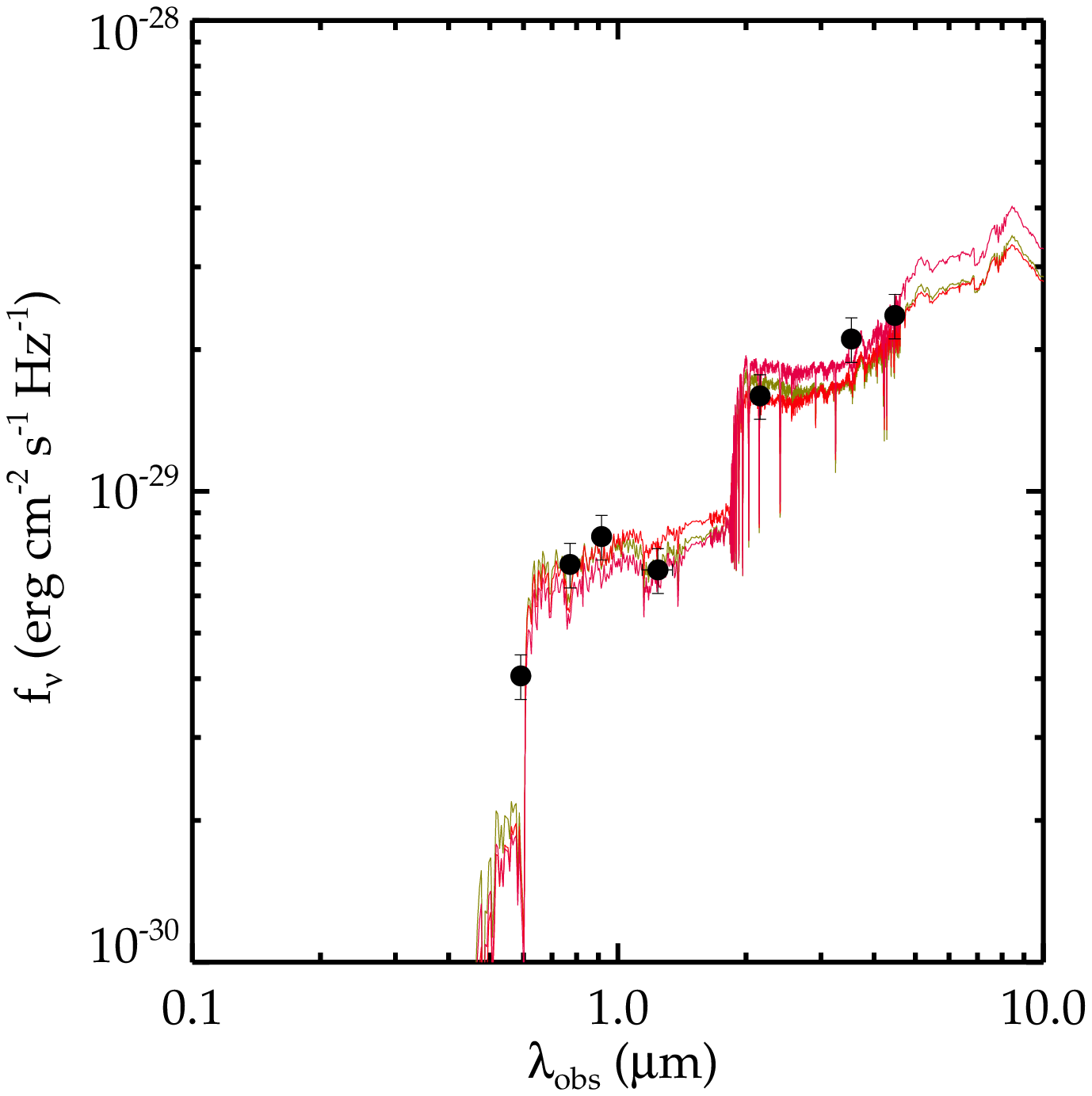}  \plotone{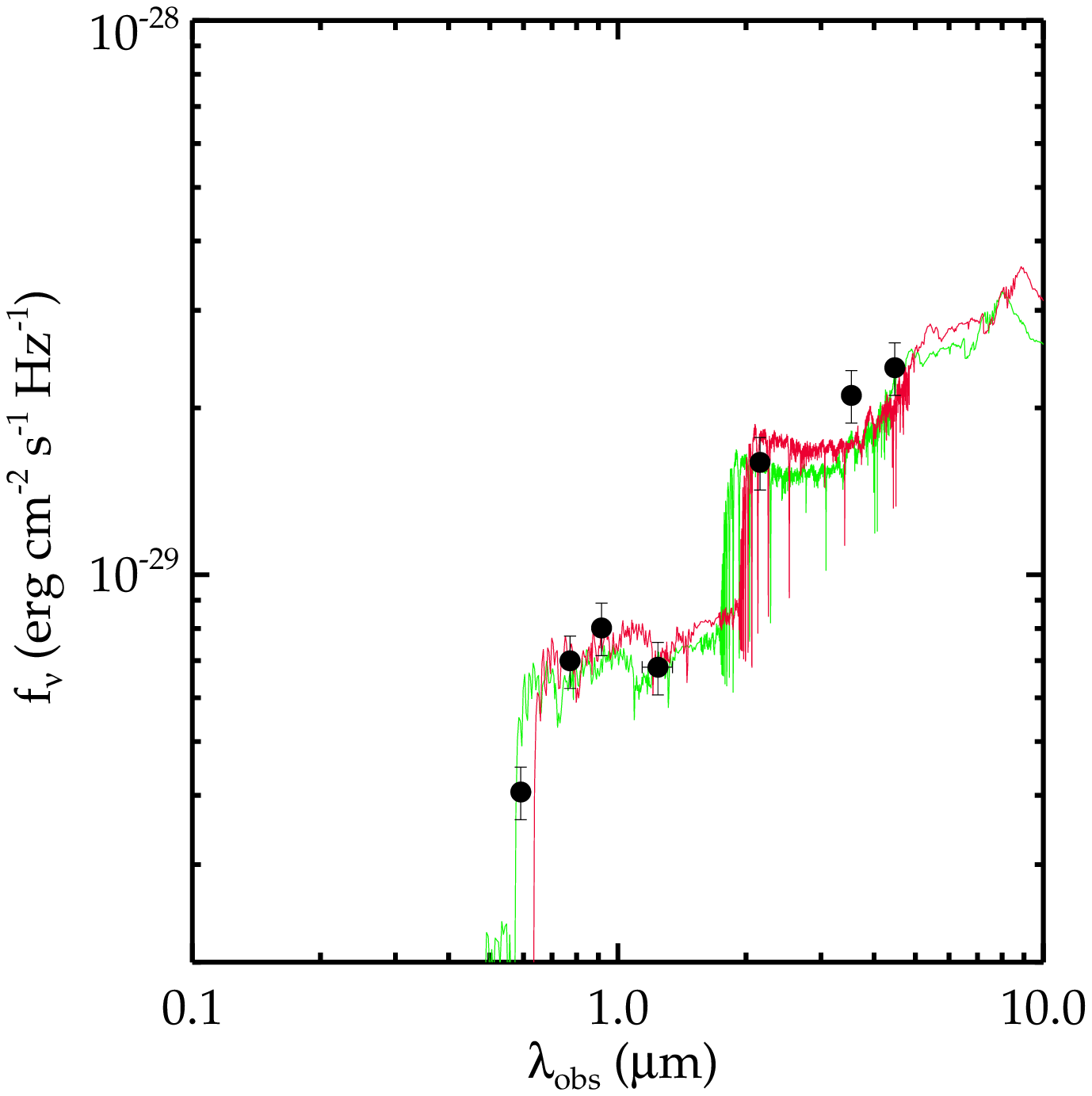}
\caption{SED of a B-drop (typical of bright sources in  our sample)
  compared with a range of CB07 model SEDs that fit the  data.  {\it
    Top:}  Adopting a model with an exponentially decaying star
  formation history with $\tau$=100 Myr, we  find the SED is best fit
  by an age of 286 Myr and a stellar mass of 2.3$\times$10$^{10}$
  M$_\odot$ with E(B-V)=0 (green curve).  Allowing the dust content to
  vary, we find  acceptable fits for stellar  masses ranging between
  2.0 and 2.8$\times$10$^{10}$ M$_\odot$ and ages between 180 and  321
  Myr. The model with the lower mass and age is shown as the red curve
  (and has E(B-V)=0.1); the model that provides the upper bound to the
  mass and age is shown in purple (and has E(B-V)=0). {\it Bottom:}
  Perturbing the photometric redshift by $\Delta z=\pm$0.25, we find
  that the  higher redshift solution (red curve) provides a
  significantly less  accurate fit to the data ($\Delta
  \chi_{reduced}^2$= 3.06) than that provided by the $\Delta z=-0.25$
  SED (green curve) and the SED derived using the best fit photo-$z$.
  The inferred stellar mass in the lower redshift model 
  (2.3$\times$10$^{10}$ M$_\odot$) is within 5\% of the best-fit value 
  at the best-fit BPZ photometric redshift and the inferred age (203 Myr) is 
  within 30\%.}
\end{figure}

Extracting robust information from the SEDs of faint dropouts is more
difficult than for the bright sources discussed above, largely because
the near and mid-infrared datasets are not deep enough to detect very
faint objects.  Sources that are undetected in IRAC have a median mass
of 2$\times$10$^{8}$ M$_\odot$ and age of 80 Myr for the $\tau$=100
Myr exponential decay model, but their SEDs have acceptable masses and
ages spanning a factor of 10.  In general, we find that in order to
achieve reasonably reliable masses for individual  objects, we must
limit our sample to those dropouts with $\gsim 10^{9.5}$ M$_\odot$.
In Figure 7, we present  the SED of a typical object near this mass
limit (best fit stellar mass of 5$\times$10$^{9}$ M$_\odot$); acceptable CB07
fits  for the $\tau$=100 Myr exponentially decay star formation
history are found  using models with stellar masses between 2.4 and
7$\times$10$^{9}$ M$_\odot$, ages between 21 Myr and 360 Myr, and
color excesses between  $E(B-V) = 0$ and $0.3$.  While the data allow a
wide range of ages and dust extinctions, the stellar mass  can be
determined reasonably well.

\begin{figure}
\figurenum{7} \epsscale{1.0}  \plotone{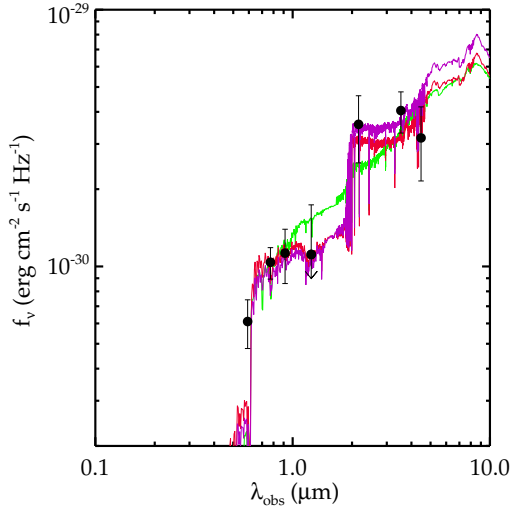}
\caption{SED of a B-drop typical of sources at the adopted 10$^{9.5}$
  M$_\odot$ stellar mass limit.  Adopting a model with an
  exponentially decaying star formation history with $\tau$=100 Myr,
  we find the SED is best fit by an age of 320 Myr and a stellar mass
  of 4.8$\times$10$^{9}$ M$_\odot$ with E(B-V)=0 (red curve).
  Allowing the dust content to vary, we find acceptable fits for
  stellar masses ranging between 2.4 and 6.7$\times$10$^{9}$
  M$_\odot$ and ages between 21 and 360 Myr. The model with the
  lower mass and age is shown as the green curve (and has E(B-V)=0.3);
  the model that provides the upper bound to the mass and age is shown
  in purple (and has E(B-V)=0). }
\end{figure}

To obtain an estimate of the average properties of galaxies less
massive than this limit, we perform a stacking analysis on the low S/N
infrared data.  For each of the three dropout samples, we select all
sources that are undetected and unconfused at 3.6 $\mu$m, amounting to
171 B-drops, 29 V-drops, and 17 $i'$-drops.  For each of the dropout
samples, the median $z'_{850}$-band magnitude of this faint subset is
$\simeq 26.7$.  Following the techniques discussed in \cite{Eyles07},
we stack the IRAC data of each of the undetected dropouts (excluding
those with $\lsim 65$ ksec of exposure time) using a median combine
with no weighting to reduce the effect of contamination by neighboring
sources.  We follow a very similar approach in stacking the near-IR
data, limiting our sample to the objects in GOODS-S given the slightly
deeper images in that field.

\begin{deluxetable}{cccc}
\tablewidth{0pc} \tabletypesize{\scriptsize}  \tablecaption{Photometry
  of Stacked Images}
\tablehead{
  \colhead{J} &  \colhead{$K_s$} &  \colhead{m$_{3.6\mu m}$} &
  \colhead{m$_{4.5\mu m}$}}  \startdata \multicolumn{4}{l}{\hrulefill
  B-drops ($z\simeq 4$) \hrulefill}\\* 27.03(0.15)  & 26.51(0.15)  &
26.79(0.31) & 27.29 (0.17) \\   \multicolumn{4}{l}{\hrulefill V-drops
  ($z\simeq 5$) \hrulefill}\\*  26.7(0.34)  & 26.55(0.48)  &
26.89(0.21) & 27.71(0.57) \\   \multicolumn{4}{l}\hrulefill
{$i'$-drops ($z\simeq 6$)\hrulefill}\\*   26.69(0.32)  & 26.43(0.42) &
27.23(0.33) & 27.64(0.87) \\   \enddata \tablecomments{The photometric
  errors of the stacked images are provided  in parenthesis next to
  the measured magnitudes.}
\end{deluxetable}

The stacking procedure described above resulted in near and mid-IR
detections for  the B, V, and $i'$-drops (see Table 1 for photometry).
The model fit to the composite faint B-drop subsample SED is displayed
in Figure 8.  For the exponential-decay ($\tau=100$ Myr) model, the
best-fit solar metallicity model has  a stellar mass of 1.7$\times$
10$^8$ M$_\odot$ (acceptable fits range between 6.2$\times$10$^{7}$
and  3.2$\times$10$^{8}$ M$_\odot$) and age of 48 Myr.  We find
similar best fitting models for the stacked V-drops
(1.5$\times$10$^{8}$ M$_\odot$ and 25 Myr) and $i'$-drops
(2.6$\times$10$^{8}$ M$_\odot$ and 38 Myr).  In all cases, the  SED is
actually slightly better fit by a sub-solar metallicity model ($\Delta
\chi^2\simeq 0.4$-0.6) in agreement with the results of
\cite{Eyles07}; however, the returned stellar masses and ages do not
change significantly from those obtained using the solar metallicity
models (Figure 8).  Importantly, the inferred masses and ages of the
stacked SEDs are consistent with  what we would have obtained by
merely taking the average of the  masses and ages of the individual
galaxies included in the stack, suggesting that in spite of the large
uncertainties, we can still use the inferred properties of this faint
subset and  be confident that the average properties are robust.

\begin{figure}
\figurenum{8} \epsscale{1.0}  \plotone{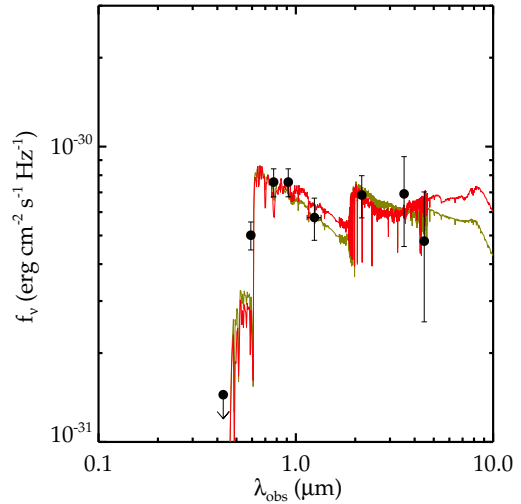}
\caption{Fit to composite SED of faint B-drop stack.  The best-fitting
  exponential decay $\tau$=100 Myr model with solar metallicity (red
  SED) has a stellar mass of 1.7$\times$ 10$^8$ M$_\odot$ and age of
  $\simeq 50$ Myr. Using a subsolar metallicity, Z=0.2$Z_\odot$,
  template (green SED), we find a marginally better fit to the data
  ($\Delta\chi^2$= 0.5) with similar inferred masses and ages
  (2.4$\times$10$^{8}$ M$_\odot$  and $\simeq 70$ Myr).}
\end{figure}

\section{New Insight into the History of High Redshift Star Formation}

One of the most basic questions in the study of galaxy formation is
how galaxies grow over cosmic time.  Observations have now  revealed
vast populations  of vigorously assembling galaxies at $z\gsim 4$, but
it remains unclear how star formation proceeds in these objects.   On
one hand it is possible that $z\gsim 4$ LBGs assemble their mass at a
constant rate, growing continuously between $z\simeq 6$ and 4.  This
``steady growth'' scenario may be expected if high redshift galaxies
are constantly fueled by cold gas which is in turn rapidly converted
into stars.   Alternatively, galaxies may undergo punctuated star
formation episodes lasting only several hundred Myr or less, such that
those LBGs seen at $z\simeq 5$, for example, are not necessarily
related to  those seen at later times (e.g., at $z\simeq 4$) or at
earlier times (e.g.,  $z\simeq 6$).  This ``independent generations''
scenario could arise if relatively rare events (e.g., major mergers,
instabilities) are predominantly responsible for sparking luminous
star formation at $z\gsim 4$, or alternatively if feedback effects act
to shut off star formation after several hundred Myr.

Testing the validity of these scenarios has proven difficult with
rest-UV observations alone, as these studies only probe recent star
formation activity, revealing little about the physical properties of
the galaxies being assembled.   But it is the evolution of such
properties (i.e., stellar mass, age, dust) which can potentially
illuminate the relationship between galaxies at $z\simeq 4$ and
$z\simeq 6$.  If the LBGs seen at $z\simeq 4$ have been forming stars
at a constant rate since $z\simeq 5-6$, then the higher redshift
objects of a given luminosity should appear as ``scaled-down''
versions of those at $z\simeq 4$, with lower masses, ages, and reduced
dust content.  In contrast, if the LBG population at $z\simeq 4$ is
dominated by galaxies which were either quiescent or forming stars
less vigorously at $z\simeq 5$ and 6, then we would not necessarily
expect  the stellar populations to show signs of growth between the
two different  epochs.  

In this section, we attempt to distinguish between these various
scenarios by exploring the evolution of the relationship between the
stellar mass and UV luminosity (hereafter the M$_\star$-M$_{1500}$
relation).  For a given galaxy, this diagnostic illuminates the ratio
of the past and present star formation.  At lower redshifts, this
relationship has been used effectively to constrain stellar mass
growth  in galaxies \citep{Noeske07,Daddi07}.  Studying the redshift
evolution of this relation for a large sample of high redshift
dropouts reveals whether the typical specific star formation rate
(SFR/M$_\star$) and stellar mass of UV luminous galaxies changes
significantly between $z\simeq 4$ and 6, allowing new constraints to
be placed on the past star formation history.  

\begin{figure}
\figurenum{9}
\begin{center}
\epsscale{1.2}
\plotone{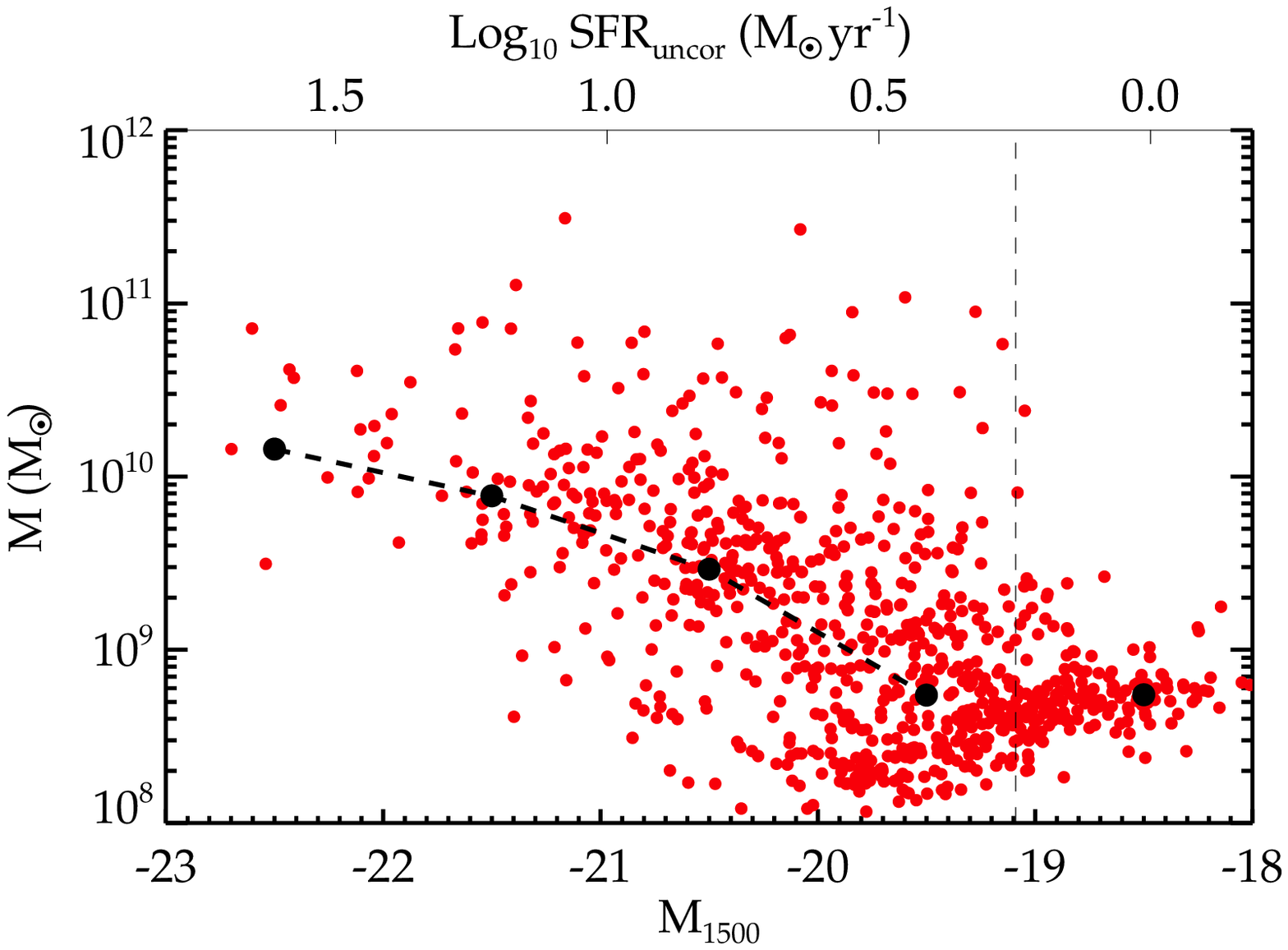}
\plotone{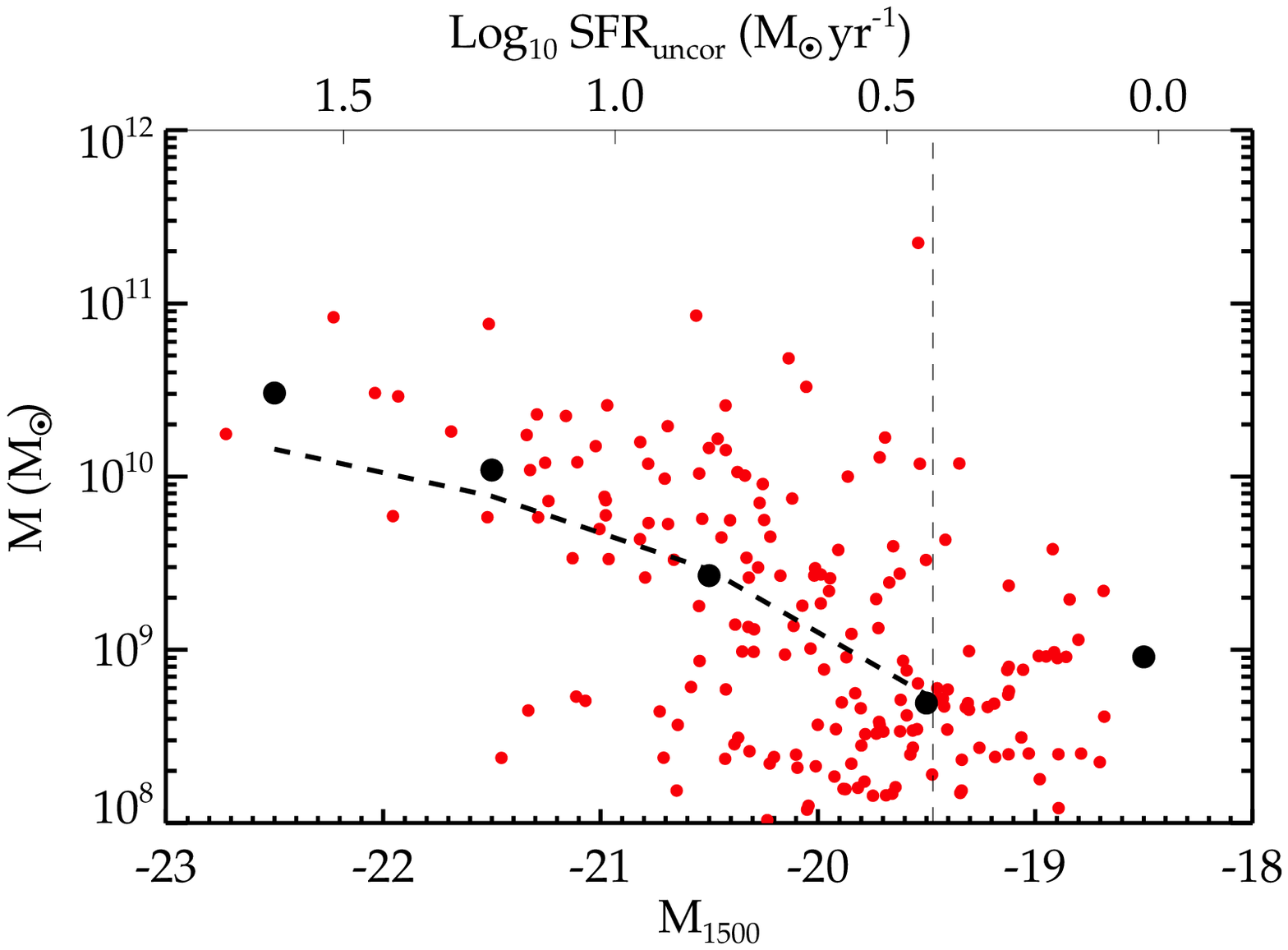}
\plotone{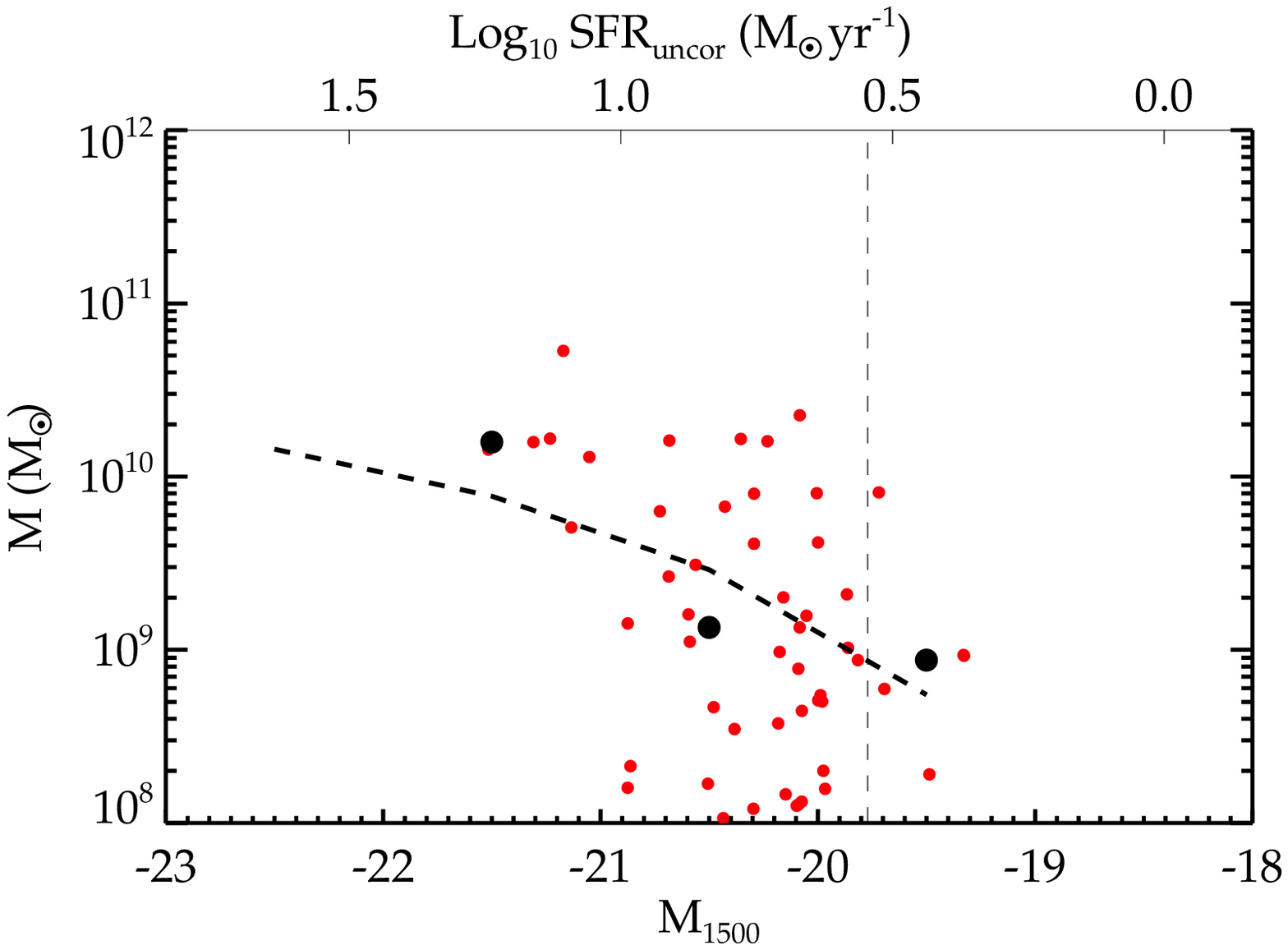}
\end{center}
\caption{Stellar mass vs. absolute magnitude at 1500\AA~(uncorrected
  for dust extinction) over $z\simeq 4-6$.  Small red circles
  correspond to inferred stellar masses and rest-UV absolute
  magnitudes for individual B-drops (top), V-drops (middle) and
  $i'$-drops  (bottom) assuming a $\tau=100$ Myr exponential decay
  model.  The dark solid circles are the median stellar mass in each
  magnitude bin. The relationship at $z\simeq 4$ is overlaid on the
  $z\simeq 5$  and $z\simeq 6$ panels as a black solid  dashed
  line. The vertical solid lines represent the adopted completeness
  limits for each sample.  The median stellar mass increases
  monotonically with M$_{1500}$ in each dropout sample; however at a
  fixed M$_{1500}$, the median stellar mass  does not decrease 
  significantly with increasing redshift, as may be expected in simple 
  steady growth models. }
\end{figure}

In Figure 9, we present the inferred stellar masses of the B, V, and
$i'$-drops as a function of absolute magnitude at rest-frame 1500~\AA~
(M$_{1500}$).  We compute the absolute magnitude from the observed
$i'$-band magnitudes for the B-drops, and the $z'$-band  magnitudes
for the V- and $i'$-drops using photometric redshifts where
spectroscopic  confirmation is not available.  Ideally we would
correct the absolute magnitudes for dust extinction, but doing so can
introduce  considerable error into the relation.  For example, using
the Meurer relation \citep{Meurer99} we find that the extinction  at
1600~\AA, A$_{1600}$,  is related to the rest-UV color for dropouts at
$z=4$ by the following relation: A$_{1600}$=0.415 +
10.9($i'_{775}$-$z_{850}$).  This indicates that photometric
uncertainties as small as 0.1 mags in the rest-UV color translate to
more than 1 mag of uncertainty in the extinction correction.  Further
error arises from uncertainty in the redshifts and, to a lesser
extent, the ages of the dropouts, both of which alter the relation
between extinction and color derived above.  Together, these sources
of error would undoubtedly add significant random scatter to the
M$_\star$-M$_{1500}$ relation, quite possibly washing out any trends
that may exist with luminosity or redshift.  We thus limit our
analysis to the relationship between stellar mass and emerging
luminosity  (i.e., the luminosity inferred from the flux that escapes
the galaxy) and comment below on the specific effects that not
correcting  for dust extinction may have on our results.

For each of the dropout samples, we find that the median stellar mass
increases monotonically with increasing emerging UV luminosity
(forming a so-called ``main-sequence''), albeit with very significant
scatter in each magnitude bin.  Similar relations are commonly  seen
at lower redshifts (e.g., \citealt{Daddi07}) and are predicted to
arise at high redshift in hydrodynamic simulations (e.g.,
\citealt{Finlator06}).  In the idealized scenario in which the bright
dropout population emerges at $z\simeq 6.5$ and steadily grows in
stellar mass via a constant star formation rate (with no new galaxies
emerging as luminous dropouts), we would expect the typical dropout
stellar masses to increase by a factor of $\simeq 7$ in the time
between $z\simeq 6$ (when the galaxy has an age of $\simeq 90$ Myr) and 
$z\simeq 4$ (when the galaxy has an age of $700$ Myr).  

\begin{deluxetable}{ccccc}
\tablewidth{0pc} \tabletypesize{\scriptsize} \tablecaption{Evolution
  of Galaxy Properties over 4$\lsim$z$\lsim$6}
\tablehead{
  \colhead{M$_{\rm{UV}}$} & \colhead{Num} & \colhead{M$_{\star}$
    ($\tau$=100)} & \colhead{M$_{\star}$ (CSF)}  & \colhead{Age} \\
  \colhead{(mag)} & \colhead{} & \colhead{(10$^8$ M$_\odot$)} &
  \colhead{(10$^8$ M$_\odot$)} & \colhead{(Myr)} } \startdata
\multicolumn{5}{l}{\hrulefill B-drops ($z\simeq 4$) \hrulefill}\\*
-22.5   & 15   & 144 (31-410)  & 223  &  180 \\  -21.5   & 73   & 77
(24-380)  & 114  &  203 \\  -20.5   & 227  & 29  (4.0-130) & 38   &
181 \\  -19.5   & 370  & 5.5 (2.0-42)  & 7.3  &   143
\\  \multicolumn{5}{l}{\hrulefill V-drops ($z\simeq 5$) \hrulefill}\\*
-22.5   & 3    & 304 (180-830) & 520   & 286  \\  -21.5   & 20   & 109
(5.4-234) & 112   & 227 \\  -20.5   & 67   & 27  (2.2-158) & 36    &
181   \\  -19.5   & \ldots   & \ldots    & \ldots   &     \ldots
\\  \multicolumn{5}{l}\hrulefill {$i'$-drops ($z\simeq 6$)
  \hrulefill}\\*	 -22.5   & 0    & \ldots         & \ldots &
\ldots \\  -21.5   & 8    & 160 (24-530)   & 226    &   286 \\  -20.5
& 41   & 13 (1.3-90)    & 27     &   102   \\  -19.5   & \ldots   &
\ldots     & \ldots &  \ldots  \\    \enddata \tablecomments{The
  stellar masses and ages are inferred from models using  a Salpeter
  IMF and solar metallicity. In column 3, we present the median
  stellar masses (and the range of masses spanned by the middle 80\%
  of the  distribution) determined from an exponentially-declining
  star formation history with $\tau$=100 Myr.  In column 4, we provide
  the median stellar masses  inferred assuming a constant star
  formation history. In column 5, we present the median ages inferred
  for the $\tau$=100 Myr models.}
\end{deluxetable}
 
Examining the top, middle, and bottom panels of Figure 9, we do not
see such a systematic increase in the normalization of the
M$_\star$-M$_{1500}$ relation between $z\simeq 6$ and $4$ (Table 2).
While the lowest luminosity bin above our completeness limit may show
some signs of marginal stellar mass growth between $z\simeq 6$ and 5
(i.e., a $\times 1.3-2.0$ increase  in the median stellar mass),  this
same bin shows no growth between $z\simeq 5$ and 4 in spite of the
improved mass  estimates and statistics available at the lower
redshifts.  The most luminous bins actually show a slight decrease in
the average stellar mass with cosmic time.  As shown in Table 2, the
absence of a systematic increase in the average stellar masses is not
strongly dependent on the chosen star formation history.  Overall
these results seem to imply that the ratio of median stellar mass to
emerging UV luminosity does not evolve significantly for LBGs over
$z=4-6$.  A galaxy with a given M$_{1500}$ at $z\simeq 6$ will, on
average, have the same assembled mass (to within a factor of $\simeq
2$) as a galaxy seen at $z\simeq 4$ with the same M$_{1500}$.  This
suggests  that the specific star formation rate evolves weakly over
$4\lsim z\lsim 6$, indicating that the typical duration of past star
formation for a galaxy of a given luminosity does not vary
significantly between $z\simeq 6$ and 4.  

While the inclusion of a dust correction would shift the
M$_\star$-M$_{1500}$ relation brightward (i.e., to the left in Figure
9), it would likely not lead to an increase in the normalization of
the M$_\star$-M$_{1500}$ relation over time.  As has been shown
elsewhere  (e.g., \citealt{Stanway05, Bouwens07}) galaxies at
M$_{1500}<-$19.8 do potentially become marginally dustier between
$z\simeq 6$ and 4  which would cause the $z\simeq 4$
M$_\star$-M$_{1500}$ relation to shift slightly more than the $z\simeq
5$ and 6 relations.  Since a relative shift brightward is roughly the
same as a shift toward lower stellar masses at fixed M$_{1500}$, this
would actually have the effect of slightly decreasing the
normalization of the M$_\star$-M$_{1500}$ relation over the $4\lsim
z\lsim 6$ redshift range.

\subsection{Testing the Steady Growth Scenario}

We now attempt to place the M$_\star$-M$_{1500}$ relation presented in
Figure 9 in the context of the steady growth scenario discussed at the
outset of this  section.  To approximate this scenario, we assume that
galaxies follow a constant star formation rate.  Using solar
metallicity templates and allowing  the dust content to freely vary,
we find that the typical implied star formation lifetimes of the
B-drops are in excess of 700 Myr in the two brightest bins (Figure
10), implying that the precursors of the majority of B-drops would
have been equally luminous in V and $i'$-drop samples.  We therefore
would have expected to see a steady decline in the typical stellar
masses and ages of UV luminous galaxies at $z\simeq 5$ and 6.  The
fact that the $z\simeq 6$ LBGs are, on average, quite similar in mass
and age to those of similar luminosity at $z\simeq 4$ suggests that
the precursors of most of the $z\simeq 4$ LBGs were not at the same UV
luminosity  (or perhaps were even quiescent) at $z\simeq 6$.  Similar
arguments are arrived at by considering the observed  UV luminosity
function (UV LF) of the dropouts over this redshift interval
\citep{Bouwens07}.  The large implied ages mentioned above combined
with the constant past star formation rate suggest that the number
density of galaxies at a given M$_{1500}$ should not fall off
significantly  with increasing redshift, in sharp contrast to the
observed trend.  Indeed, the distribution of ages implied by this
constant star formation rate scenario is such that the number density
of galaxies at M$_{1500}<-21.5$  should not decrease by more than a
factor  of $\simeq 2$ over $4\lsim z\lsim 6$, in marked contrast to
the observed $\times$10 decline in the bright end of the luminosity
function at M$_{1500}\lsim -21.5$ over this redshift interval
\citep{Bouwens07}.  

We therefore argue that in order to explain the constancy in the
M$_\star$-M$_{1500}$ relationship and the decline in the UV luminosity
function, we require a star formation model that implies that a
significant fraction of UV-luminous galaxies at $z\gsim 4$ emerged at
their present UV luminosity recently enough such that they would not
have been visible as dropouts at the same UV luminosity just $\simeq
300$ Myr earlier (e.g., the time corresponding to  $\Delta z\simeq 1$
at $4\lsim z\lsim 6$).  With a constantly increasing flux of newly
luminous galaxies, the galaxy population at each epoch would likely be
dominated by relatively young ($\lsim 300$ Myr) objects.  Hence the
main sequence of the M$_\star$-M$_{1500}$ relation would not
necessarily grow significantly with cosmic time as one would expect in
the steady growth scenario. 

We note that this requirement of a young population is not inherently
inconsistent with constant, sustained  star formation in galaxies.
Indeed, if each of the $z\gsim 4$ dropout populations is dominated in
number  by newly emerged luminous galaxies, it is possible that
systems which are UV luminous at  $z\simeq 6$ remain luminous to
$z\simeq 4$, but make up such a small fraction (by number) of the
$z\simeq 4$  LBG population so as not to affect the main sequence of
the M$_\star$-M$_{1500}$ relation.  The key problem with this picture,
as hinted by our results above,  is that constant star formation
models struggle to produce young enough ages to be consistent with the
need to have each epoch (i.e. $z\simeq 4$) dominated by systems that
weren't present at the previous epoch (i.e.  $z\simeq 5$).  

\begin{figure}
\figurenum{10}
\epsscale{1.25}
\plotone{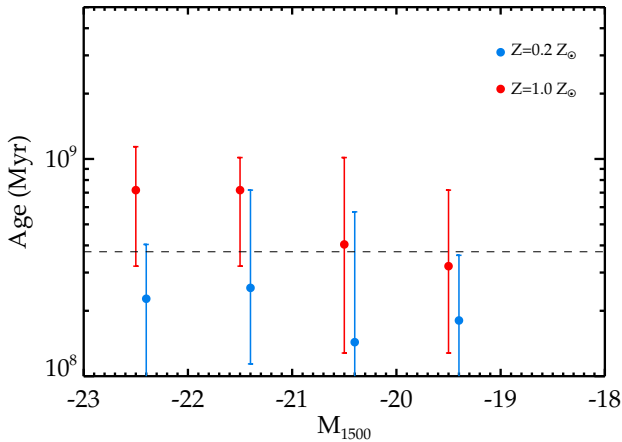}
\caption{Ages inferred for B-dropout sample from constant star
  formation scenario as a function  of absolute magnitude at
  1500~\AA~(uncorrected for dust).  The solid red  circles give the
  median inferred age for CB07 models with solar metallicity  while
  the blue circles give the median ages for models with 20\% solar
  metallicity.  The error bars on each datapoint correspond to the
  25th  (lower bar) and 75th (upper bar) percentile in the
  distribution of ages  in each absolute magnitude bin.  The
  horizontal dashed line denotes  the amount of time spanned between
  $z\simeq 4$ and 5.  Thus if the inferred ages are above the dashed
  line, it indicates a large fraction of the B-drops began  forming
  stars beyond $z\simeq 5$, implying an abundant precursor  population
  with scaled-down masses relative to those measured at  $z\simeq 4$.
  The  lack of evolution in the normalization of the
  M$_\star$-M$_{1500}$ relationship over $z\simeq 4-6$ (Figure 9)
  casts doubt on this scenario.  }
\end{figure}

It is valuable to consider the systematic uncertainties that could
arise  following degeneracy in derving both age and dust extinction
measures from  our photometric data.  If our dust estimates are
systematically too low, then the ages quoted above are likely
overestimates.  In addition, if we adopt lower metallicity templates
in our models (e.g., 0.2  Z$_\odot$) as may be appropriate for $z\gsim
4$, we find ages that are a factor of $\simeq 2$ lower than those in
Table 2 (Figure 10).  Thus while the relatively large ages implied by
the constant star formation rate scenario do cause some tension with
the observed evolution in the M$_\star$-M$_{1500}$ relation and UV LF,
there are feasible steady growth scenarios that are potentially
consistent with the observations.  Improved spectroscopic completeness
and infrared photometry will soon enable us to lessen the uncertainty
in the dust content and to provide estimates of the metallicities,
both of which will  help test the feasibility of these scenarios.

A possible variant on the ``steady growth'' scenario is a situation
where star formation grows increasingly vigorous as galaxies grow in
mass at  early times.  While such star formation histories are not
typically used in fitting high redshift galaxies, they are predicted
in SPH simulations \citep{Finlator06} and have successfully been used
to infer the properties of two $z\gsim 6$ galaxies \citep{Finlator07}.
In this case, galaxies would grow along the ``main sequence'' of the
M$_\star$-M$_{1500}$ relation of Figure 9, which results in little
evolution of the normalization of the M$_\star$-M$_{1500}$ relation
over 4$\lsim z\lsim 6$, similar to the observed trend.  This picture
is also able to explain the observed luminosity function evolution.
If galaxies in the luminosity range probed by our study steadily
brighten by $\simeq 0.7$ mag at 1500~\AA~ in the $\simeq 620$ Myr
between $z\simeq 6$ and 4 (matching the evolution  seen in the
characteristic UV luminosity, M$^\star_{\rm{UV}}$), then this scenario would be
fully consistent with the observed evolution in the  UV luminosity
function.  

Importantly, a significant obstacle for each of the ``steady growth''
scenarios described above comes from observations  of the clustering
of dropouts at $z\gsim 4$ \citep{Ouchi04,Lee06,Lee08}.  By linking the
dropouts of a given UV luminosity to a dark matter halo mass,
constraints can be placed on the fraction of halos which are
UV-bright, from which a star formation duty cycle can be determined.
Using this approach, \cite{Lee08} concluded that the typical duration
of star formation activity in the dropouts at $z\simeq 4$ must last no
longer than $\simeq 400$ Myr.  Hence according  to these results,
those  galaxies that are active at $z\simeq 6$ would no longer be
active  at $z\simeq 4$, in direct conflict with the sustained activity
inherent to this picture.  

In summary, we find that it is difficult to reconcile the observed
data (UV LF, stellar populations, and clustering) with a scenario in
which the galaxies observed at $z\simeq 4$ have been steadily growing
at a fixed UV luminosity since $z\simeq 6$.  Instead the data require
$z\gsim 4$ dropout galaxies to have been less luminous several hundred
million years prior to the redshift at which they are observed.  Hence
we find that a promising means of reconciling the steady growth model
with the observations could be a scenario whereby the star formation
rates of galaxies are, on average, growing more vigorous as their mass
grows, as proposed by \cite{Finlator06}.

\subsection{Testing the Independent Generations Model}
   
We now consider the feasibility of the ``independent generations''
model mentioned  at the outset of this section.  In this picture, the
galaxies we  observe at $z\gsim 4$ undergo shorter timescale star
formation episodes lasting no more than $\simeq 300$-400 Myr.  In
order to achieve these relatively punctuated episodes of luminous
activity, each star formation event must eventually decline.  We model
this behavior using the simple exponential decay models discussed in
\S5.3.  The typical inferred ages for B-drops using solar metallicity
models with decay constants of $\tau$=100 (300) Myr range from 140
(230) to 200 (450)  Myr at M$_{1500}<-$19.5, significantly lower than
the constant star  formation models discussed above.  These ages are
comparable to the time  between $z\simeq 4$ and 5, indicating that
with these assumptions the $z\simeq 4$ dropout population is dominated
by sources which were not actively forming stars at $z\simeq 5$, as
required based on our discussion above.  Unless the cold gas
reservoirs present in newly emerged galaxies at a given UV luminosity
vary strongly between $z\simeq 4$ and 6, we would expect to see very
little evolution in M$_\star$-M$_{1500}$ relation, similar to the
observations  presented in Figure 9.  

Since the inferred ages of the dropouts in this episodic star
formation scenario are typically small enough such that they would not
have been active in higher redshift samples, there is no inconsistency
with the observed decrease in the number density of luminous sources
at $z\gsim 4$.  The only requirement is that whatever process
initiates the star formation activity must steadily produce more
vigorous episodes over cosmic time.  Given the expected growth in the
underlying dark matter halos hosting the dropouts, this is not an
unreasonable requirement. 

Unlike the ``steady growth'' scenario discussed previously, the
exponentially declining models predict that the dropouts will grow
progressively fainter over time.  Assuming models with decay constants
of  100 Myr, the sources will become nearly 3 magnitudes fainter in
M$_{1500}$ in just 300 Myr, and thus would almost certainly escape
selection in future dropout samples with similar magnitude limits
unless new star formation episodes  have been initiated.  The short
($\lsim 500$ Myr) star formation timescales implied by this scenario
are clearly in much better agreement  with those inferred from the
clustering analysis previously discussed \citep{Lee08}. 

A potential drawback of the exponentially decaying models discussed
above  is that in order to reproduce the observed SEDs of the dropout
populations in the required short timescales,  we require the past UV
luminosities to exceed those measured presently.  For the $\tau$=100
Myr models, we would expect the typical galaxy to have been a factor
of $\simeq 3$ brighter  at 1500~\AA~ just 150 Myr earlier (with this
very luminous phase lasting for a duration of $\simeq 50$ Myr).  This
implies an abundant, precursor population of very young, luminous
objects,  which are seemingly not present in sufficient number in the
observed dropout sample.  This problem could potentially be explained
if this luminous phase is generally heavily extincted by dust.
Indeed, \cite{Shapley01} found that the most luminous $z\simeq 3$ LBGs
tend to be very young and dusty.  Further spectroscopy  and higher S/N
near-IR data will soon help test this possibility.

One prediction from this episodic star formation model is that there
should be a significant repository of stellar mass in objects which
underwent a UV luminous  phase at an earlier epoch but are currently
not actively forming stars.  A careful comparison of rest-UV and
rest-optical selected samples at $z\simeq 4$ could thus yield
significant insight into the mode of star formation that is dominant
at $z\gsim 4$.  By examining the range of  star formation rates
present in galaxies of fixed stellar mass, we should  be able to
distinguish between the steady growth and independent generations
scenarios.  The reliability of this test relies on having samples with
secure redshifts and accurate extinction corrections, requiring both
additional spectroscopy and improved IR photometry.  

Before concluding this section, we must comment on several caveats
that apply to our discussion.   The first is the possibility of an
evolving initial mass function \citep{Dave08,vanDokkum08} in which the
characteristic stellar mass increases  toward higher redshifts.  As
explained in detail in \cite{Dave08}, this would  alter the current
star formation rates and the inferred  stellar masses in such a way to
decrease the normalization of the M$_\star$-M$_{1500}$ relation at
progressively earlier times.  Clearly this does not help reconcile the
observations with the steady growth picture.  The second caveat is our
neglect of the contribution of galaxy mergers to stellar mass growth.
While minor mergers are expected to be very common at high redshift,
they are likely to be gas-rich interactions.  Such merger events would
help provide galaxies with the gas required for star formation;
however, the established stellar mass gained in a gas-rich merger
would likely not be significant compared to the stellar mass growth
that is occuring due to star formation.  This view is supported by
simulations which predict that stellar mass growth is dominated by star
formation and not mergers at the high redshifts we are considering
\citep{Keres05}. 

In summary, the episodic star formation model (approximated by a
single  component exponentially declining star formation history with
luminous lifetimes  of 300-500 Myr) can  explain the evolution in the
M$_\star$-M$_{1500}$ relation, the UV LF, and the clustering
properties at $z\gsim 4$.  If this mode of star formation is dominant
at $z\gsim 4$, it would signal a potential shift from the  mode
thought dominant at $z\simeq 0.5-2$, where the lack of significant
scatter in the dust-corrected SFRs of galaxies at fixed stellar mass
\citep{Noeske07,Daddi07} argues against the feasibility of star
formation being  dominated by short-timescale episodes \citep{Dave08}.
However, given the minimal spectroscopic completeness and  low S/N of
the IR data at $z\gsim 4$, it is too premature to rule out the
sustained mode of star formation for $z\gsim 4$ galaxies.  With new IR
imaging and extensive spectroscopic surveys for high redshift dropouts
currently underway, it will soon be possible to improve the fidelity
of the results presented here as well as to robustly  test several of
the additional predictions of the episodic star formation mode
described above, both of which should lead to further progress in our
understanding of the history of star formation at $z\gsim 4$.

\section{Connecting the $z\gsim 4$ Mass Assembly History to 
Quiescent Galaxies at $z\simeq 2$}

The assembly history of massive galaxies at high-redshift ($z\gsim 2$)
has been a  very active area of research in recent years.
Near-infrared  photometry of rest-UV selected objects at $z\simeq 2-3$
has revealed  a significant population of massive galaxies at high
redshift  \citep{Shapley01,Shapley05}.  In addition, near-infrared
color selected samples have proven very effective at isolating large
samples of DRGs (e.g., \citealt{Franx03}).  The red colors  of these
objects likely arise from evolved stellar populations or dust  (or
some combination of the two).  The addition of {\it Spitzer}
photometry  (e.g., \citealt{Labbe05,Webb06,Papovich06,Reddy06}) and
very deep spectra \citep{vanDokkum04,Kriek06} has enabled the
passively evolving subset of DRGs to be separated  from the dusty
subset, providing more robust constraints on the physical properties
and evolution of massive $\gsim 10^{11}$ M$_\odot$ galaxies at $z\gsim
2$.  These studies have allowed much more accurate constraints to be
placed on the epoch of formation of the population of massive galaxies
at $z\simeq 2.3$.  \cite{Kriek06} find that the near-infrared spectra
of a small sample of $z\gsim 2.3$ DRGs are best fit by ages of $\lsim
1.3-1.4$ Gyr, suggesting that most of these systems started forming
their stars between $z\simeq 3$ and 4.   

While the formation epoch of the massive galaxies at high redshift is
becoming increasingly better defined,  it is still unclear what are
the precursors of these massive systems.  Sub-millimeter galaxies
\citep{Smail98, Chapman03, Capak08, Daddi08} are convincing candidates since
their intense star formation rates  allow a substantial amount of mass
to be assembled in a short period of time (see \citealt{Cimatti04,
  McCarthy04}).  But sub-mm sources may not provide a complete sample
of  the precursors to high redshift massive and passive systems.
Indeed,  \cite{Shapley05} have shown that for reasonable assumptions
the most massive objects in $z\simeq 2$ rest-UV selected samples are
likely to evolve into quiescent systems by $z\simeq 1.7$, with space
densities that are virtually identical to those of the passive
population at that redshift \citep{Cimatti04}.  Objects may of course
spend portions of their assembly history in both a SMG and LBG phase.
Our goal in this section is to extend the efforts of \cite{Shapley05}
to quantify the fraction of massive galaxies at $z\simeq 2-3$ that
form a significant portion of their mass in an LBG phase at $z\gsim
3.5$ and may be missed by current sub-mm galaxy surveys.

We thus set forth to compute the  number densities of massive objects
in our B, V, and $i'$-drop samples.   To derive the stellar mass
functions, we group the dropouts by stellar mass in bins of $\Delta
\rm{log_{10}M}$=1.0, spanning between 10$^{7.5}$ and 10$^{11.5}$
M$_\odot$ and then compute the number density of  dropouts in each
mass bin, accounting for the variation of the effective volume with
apparent magnitude.  We also apply a correction  to account for the
galaxies which are not included in the SED-fitting  analysis due to
being blended with foreground sources in the IRAC images.   As we
discussed in \S5.1 (see also Figure 1), the fraction of sources that
are blended varies  with $i'_{775}$ or $z'_{850}$ apparent magnitude.
We account for this slight bias by applying a magnitude-dependent
correction: after computing  the number density of galaxies in a given
stellar mass and apparent magnitude bin, we correct this density by
dividing by the fraction of high-redshift objects in that magnitude
bin which were included in the SED-fitting analysis.  In order to
provide a consistent comparison of the mass function over $z=4-6$, we
impose  an absolute magnitude limit of $\rm{M_{1500}}=-20$.  This ensures
that each dropout sample is sufficiently complete to make a reliable
measurement (see Figure 9). 

\begin{figure}
\figurenum{11} \epsscale{1.15}  \plotone{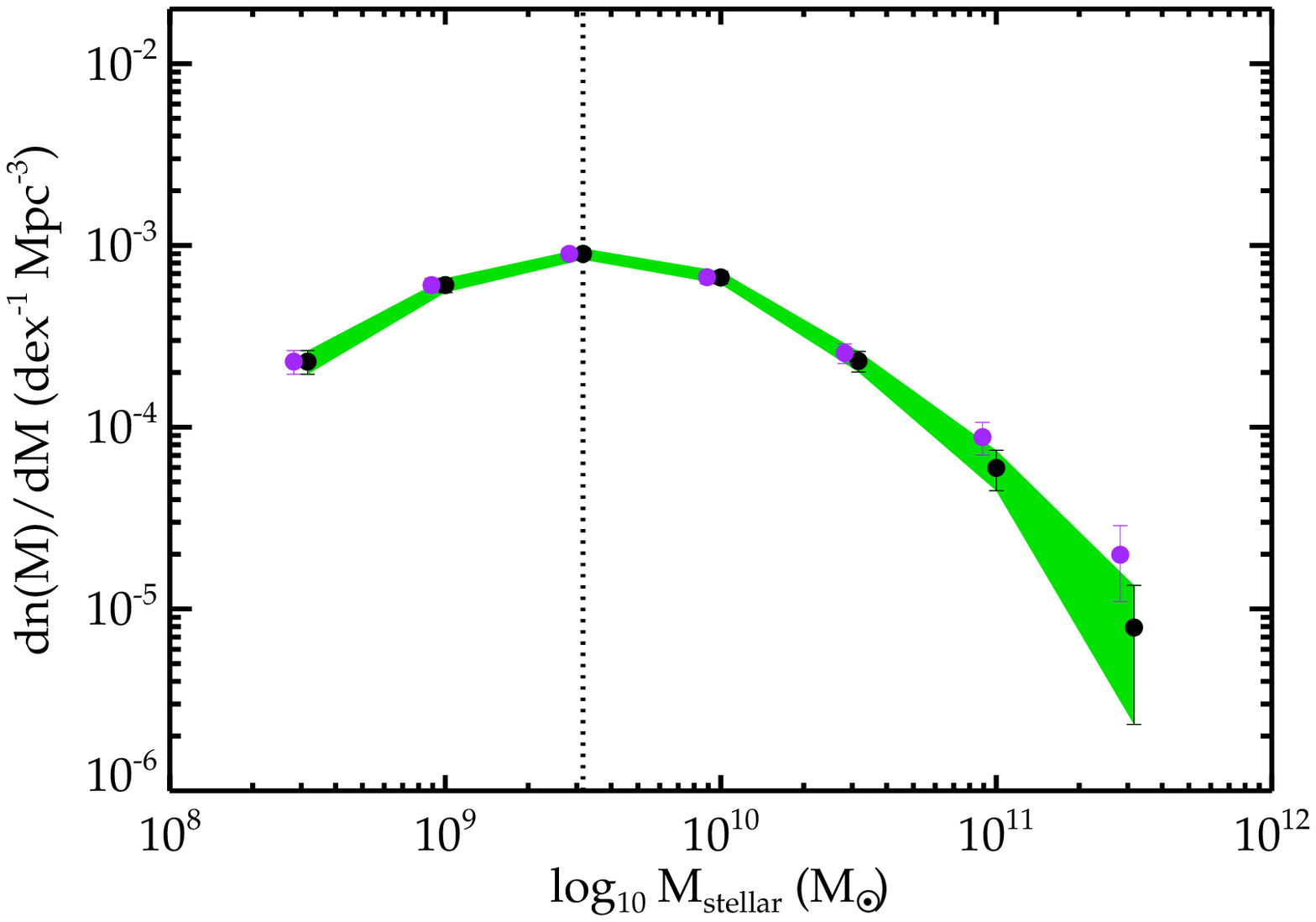} \plotone{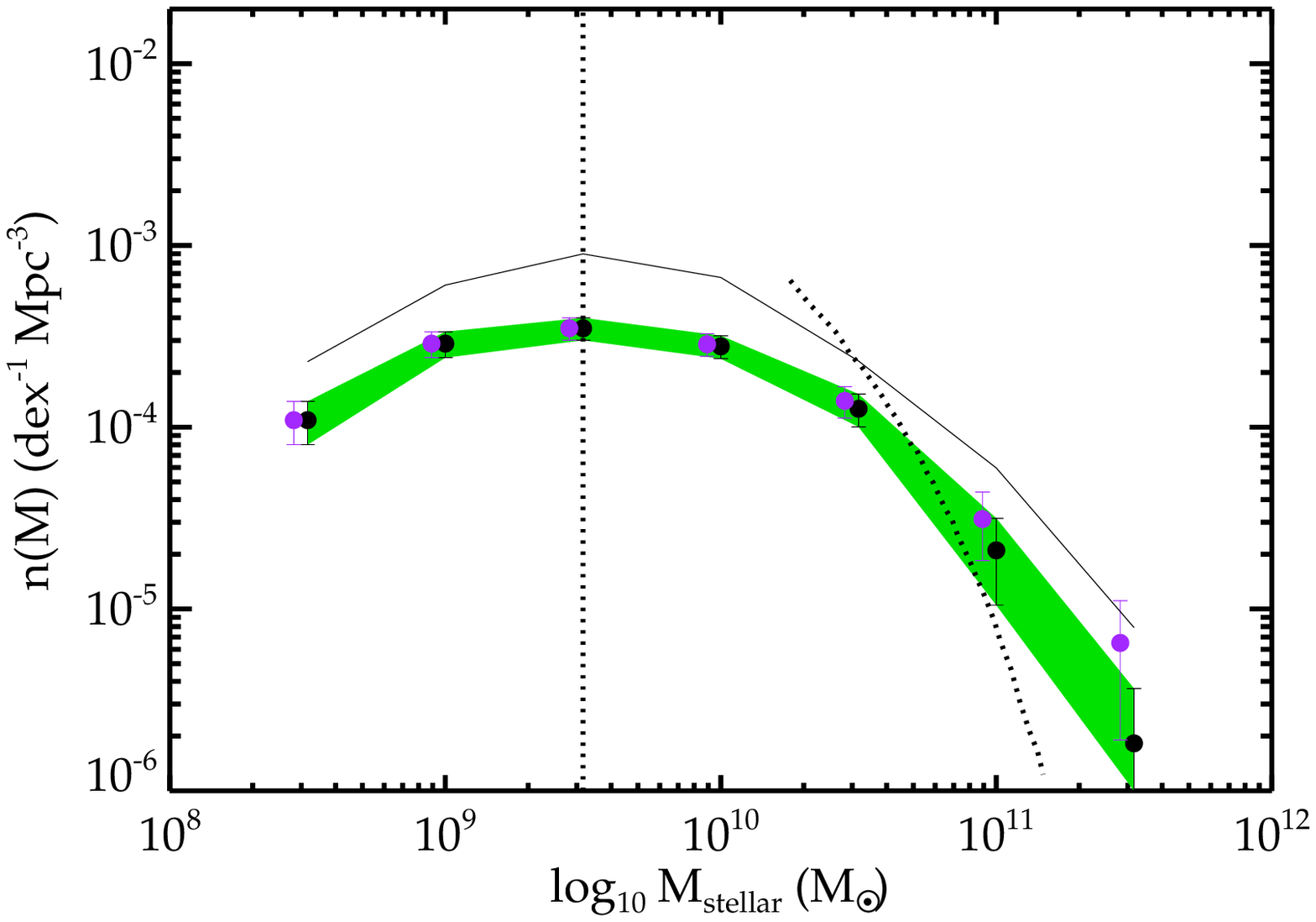}
\plotone{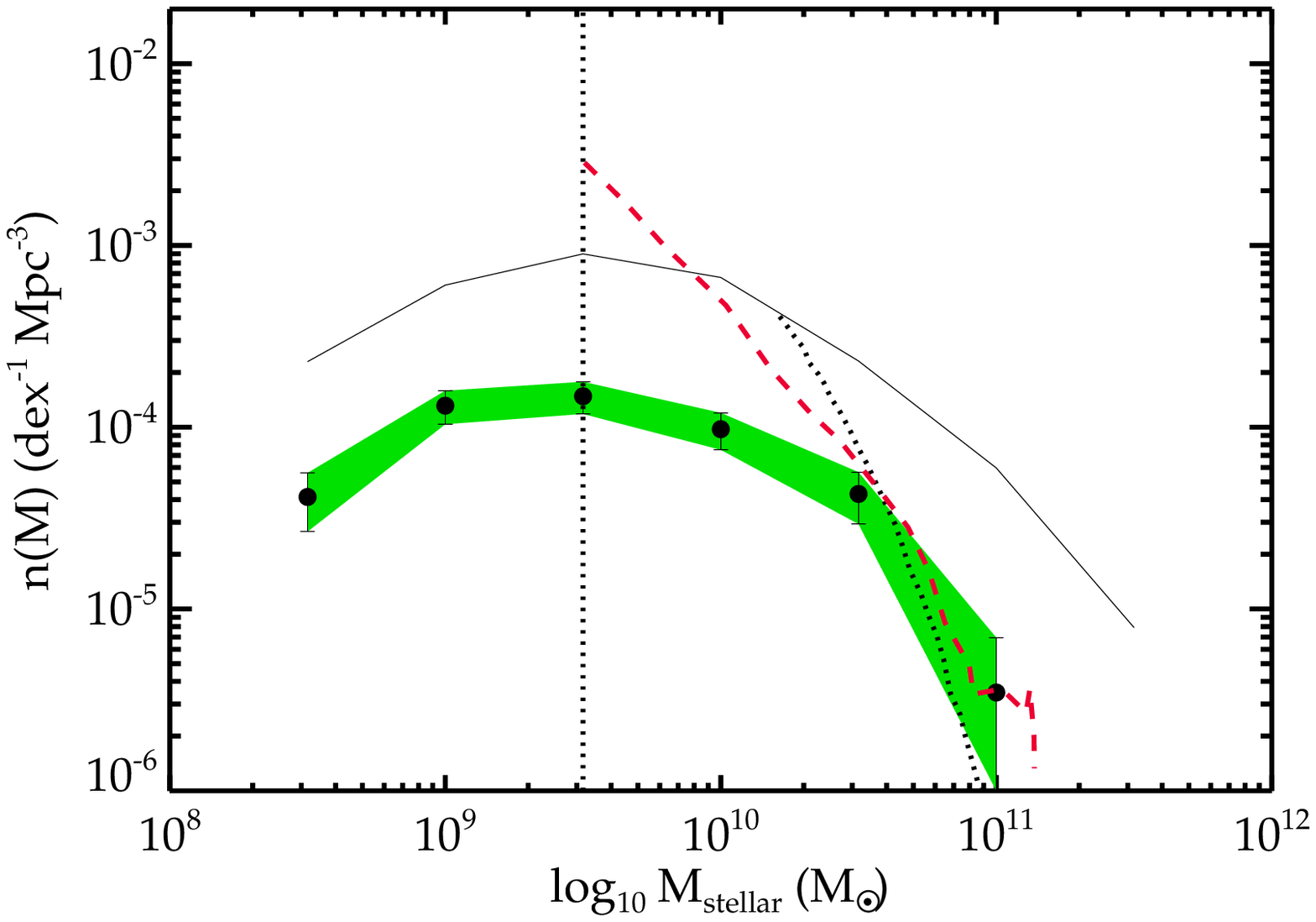}
\caption{Stellar Mass Function of B-drops (top panel), V-drops
  (middle panel), and $i'$-drops (bottom panel) with
  $M_{1500}<-20$.  The green swath and associated solid black data
  points corresponds to the mass function determined from a large
  sample of dropouts in GOODS-S and GOODS-N.  As  discussed in \S5.1,
  stellar masses are determined for individual galaxies  by fitting
  the data against CB07 population synthesis models. This plot is
  constructed from models with exponentially-decaying star formation
  histories  with $\tau$=100 Myr, solar metallicity, and a Salpeter
  IMF.  The purple datapoints (slightly offset from the central mass
  bin for clarity) correspond to the densities inferred if
  24$\mu$m-detected sources are included.  The vertical dotted lines
  represent the mass limit above  which the inferred masses are
  reliable for individual galaxies.  The  black dotted lines overlaid
  on the $V$- and $i'$-dropout mass functions (upper right and  bottom
  panels) correspond to the stellar mass function estimated in
  \cite{McLure08} by multiplying the luminosity function by the
  average M/L ratio of the dropout population.  The red dashed line
  overlaid on the $i'$-drop sample (bottom  panel) corresponds  to the
  predictions of the $z\simeq 6$ stellar mass function from the
  cosmological  SPH simulations presented in Nagamine et al. (2008).
  The black solid curves overlaid  in the upper right and bottom panel show
  the mass function determined in the  B-drop sample (upper left panel).}
\end{figure}

The resulting mass functions are shown in Figure 11.  Also shown on
the plot (purple datapoints) are the inferred  densities if we include
sources detected at 24$\mu$m in the mass function analysis.   As
expected, the 24$\mu$m detections slightly increase the number
densities of the B and V-drops in the most massive bins, but generally
do not increase them by more than the Poisson uncertainties in that
bin.  As discussed in \S4.2, the inferred stellar mass uncertainties
for individual galaxies become considerable for fainter sources with
masses  $\lsim 10^{9.5}$ M$_\odot$.  Although the average properties
of sources  less massive than this limit are reliable, we nevertheless
choose to focus our analysis on the more robustly determined objects
contained in bins 10$^{9.5}$ M$_\odot$ and above.  

The number densities derived here agree reasonably well with previous
measurements.  We plot the recent \cite{McLure08} estimate   of the
$z\simeq 5$ and 6 LBG stellar mass functions over our curves in Figure
11 (after adjusting their masses to be consistent with a Salpeter
IMF).  In spite of their approximate way of determining the stellar
mass function (the McLure et al. mass function is derived by scaling
the luminosity function by the  average stellar mass to light ratio),
the two estimates agree to within a factor of 2  over the mass range
probed, with our determination predicting a larger density of the most
massive galaxies.  The results also are in  reasonable agreement with
the massive end of the stellar mass function determined from the
cosmological SPH simulations discussed in \cite{Nagamine08}.  The
observed mass function becomes shallower  than the model predictions
at $\lsim 10^{10.0}$ M$_\odot$.  This  may be due, in part, to
observational incompleteness at low mass limits. However a similar
``knee'' in the galaxy mass function is seen in the \cite{Drory05}
data suggesting that the simulations may be underpredicting low-mass
galaxies, as suggested by \cite{Nagamine08}.

In contrast to the weakly evolving mass function at $z\lsim 1$, the
increase in stellar mass between  $z\simeq 6$ and $4$ occurs over the
entire stellar mass range considered.  Between $z\simeq 6$ and
$z\simeq 4$, we find a factor of 5-6 increase in density for the bins
ranging  between 10$^{9.5}$ and  10$^{10.5}$ M$_\odot$.  There is some
evidence that the number density grows more rapidly for the most
massive galaxies.  While the uncertainties are  significant, the mass
functions suggest that the number density  in the bin centered at
10$^{11}$ M$_\odot$ increases  by a factor of 17 between $z=6$ and
$z=4$, nearly three times greater than the growth in less massive
bins.  It seems likely that we are witnessing a  formative period of
massive galaxy formation.

\begin{figure}
\figurenum{12} \epsscale{1.2}  \plotone{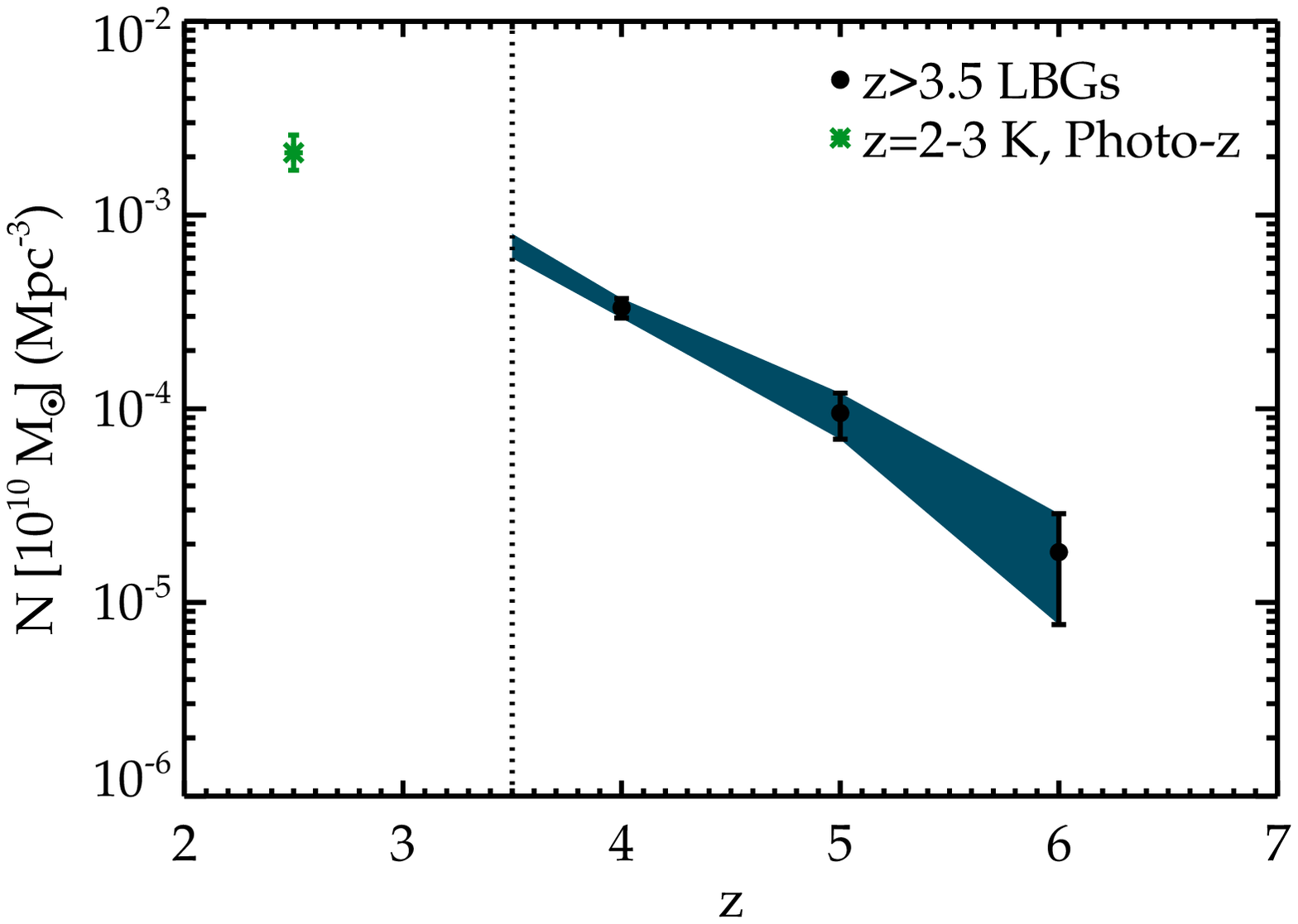} \plotone{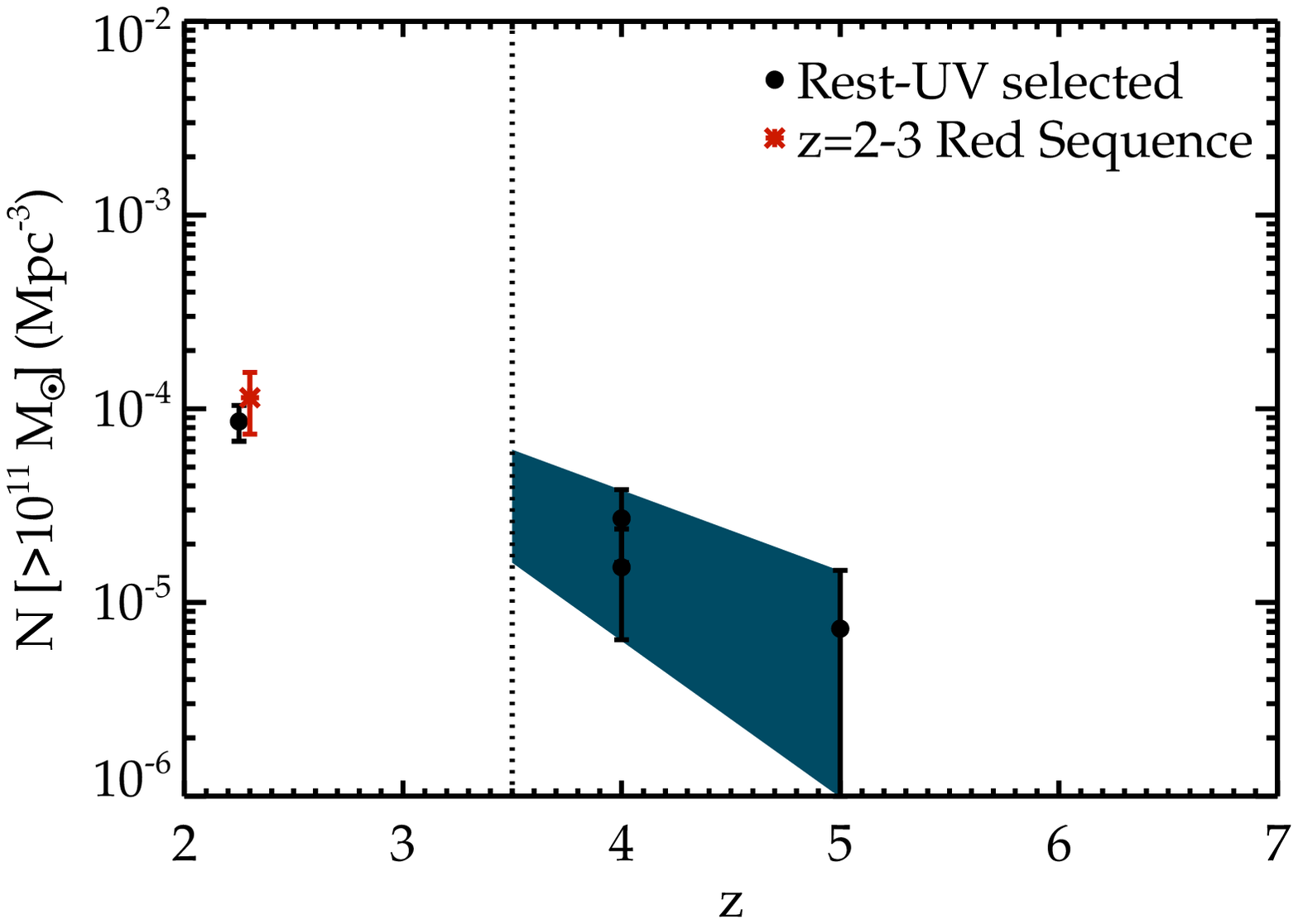}
\caption{Evolving number density of massive galaxies at high redshift.
  {\it Top:}  Solid black circles (and associated blue swath) show the
  evolution in the number  density of actively forming galaxies more
  massive  than 10$^{10}$ M$_\odot$ derived in this paper.  The green
  asterisk  correspond to the total number density of galaxies more
  massive than  10$^{10}$ M$_\odot$ at $z=2.5$ identified in a
  K$_s$-band selected  survey \citep{Drory05}.  The vertical dotted
  line demarcates the redshift range covered in this paper.  {\it
    Bottom:} Same as top panel but for galaxies more massive  than
  10$^{11}$ M$_\odot$.  The black circles give the number density of
  rest-UV selected galaxies more massive than 10$^{11}$ M$_\odot$; the
  upper datapoint at $z=4$ corresponds to the number density inferred
  if the B-drops with 24$\mu$m detections are truly at high redshift.
  The red asterisk shows the number density of  galaxies more massive
  than 10$^{11}$ M$_\odot$ that are quiescent at $z=2.3$
  \citep{Kriek08}.  }   
\end{figure}

We finally turn to whether the observed number densities of massive
LBGs at $z\gsim 2$ can account for a significant component of the
largely  quiescent population of DRGs at $z\simeq 2$.  At $z\gsim 4$,
we find that  the number density of dropouts with stellar mass greater than
10$^{11}$ M$_\odot$ (excluding MIPS-detected  sources) is $\simeq
2\times10^{-5}$ Mpc$^{-3}$ (Figure 12).  We find no overlap between
these  sources and the massive, sub-mm detected objects recently
discovered in  GOODS-N \citep{Daddi08}.  If these massive LBGs become
quiescent  on a timescale of less than 1 Gyr, then we would expect
them to appear as red galaxies with little star formation at $z\simeq
2$.  Indeed if we evolve  the colors and rest-UV luminosities of these
massive LBGs forward in  time to $z\simeq 2$ assuming the simple
exponential decay models considered  in \S6, we derive colors and
optical apparent magnitudes that are consistent  with the observed
properties of DRGs \citep{vanDokkum06}.  Comparing the  observed
density of massive LBGs at $z\gsim 4$ to the number density of
quiescent 10$^{11}$ M$_\odot$ systems at $z\simeq 2-3$
(1.1$\times$10$^{-4}$  Mpc$^{-3}$, \citealt{Kriek08}) indicates that
at least $\simeq 20$\% of the $z\simeq 2-3$ population of quiescent
10$^{11}$ M$_\odot$ galaxies formed a large fraction of their stellar
mass at $z\gsim 3.5$.  

Given the expectation that dust-enshrouded ``bursts'' of star
formation  seen in the sub-mm are largely responsible for assembling
massive  galaxies by $z\simeq 2-3$ \citep{Cimatti04,McCarthy04}, it is
interesting to extend our estimates of the  number density of massive
LBGs down to $z\simeq 2$ in order to compute the total fraction of
massive galaxies that were formed through objects present in rest-UV
selected samples.  \cite{Shapley05} estimate that the number density
of UV luminous galaxies with masses in excess of 10$^{11}$ M$_\odot$
at $z\simeq 2$ is $\simeq $10$^{-4}$ Mpc$^{-3}$.  We make our own
estimate  by constructing a sample of rest-UV selected ($z'_{850}\lsim
26.5$) galaxies  in GOODS-S with photometric redshfits between $z=2$
and 2.4 using redshifts and photometry from the GOODS-MUSIC $z'_{850}$-band
selected catalog \citep{Grazian06}.  We follow  the same procedures
described earlier for computing stellar masses, completeness as a
function of magnitude, and effective volumes.  The  derived number
densities are corrected for incompleteness (the sample is roughly 90\%
complete at $z'_{850}=26$), resulting in final number density estimates of
$\simeq \rm{9}\times\rm{10^{-5}}$ Mpc$^{-3}$, very close to the
measurement from \cite{Shapley05}.  These densities suggest that at
$z\simeq 2$, the number density of UV-bright massive galaxies is
roughly comparable to the density of systems that have already become
quiescent.  

Computing the contribution of rest-UV selected $z\gsim 2$ galaxies to
the density of quiescent $z\simeq 2$ systems is non-trivial as it
depends strongly on the timescale over which star formation is
quenched in the LBGs.  The shorter the timescale, the  larger the
implied number density of quiescent objects at $z\simeq 2$.  If we
assume that massive LBGs join the red sequence over a period of 500
Myr, then we can estimate a quiescent number  density by integrating
over the growth in UV luminous massive galaxies  over $2\lsim z\lsim
6$ (Figure 12).  Following this approach, we find that  rest-UV
selected galaxies would produce a quiescent galaxy density of
7$\times$10$^{-5}$ Mpc$^{-3}$ by $z\simeq 2.5$, increasing to
2$\times$10$^{-4}$  Mpc$^{-3}$ at $z\simeq 2$.   If the quenching
timescale is longer (i.e., 1 Gyr), then  the implied quiescent number
density decreases to $\simeq 5\times10^{-5}$  Mpc$^{-3}$ at $z\simeq
2$.  Hence, these simple assumptions suggest that star formation in
rest-UV selected $z\gsim 2$ galaxies accounts for the  assembly of at
least 50\% of the total quiescent population of massive galaxies at
$z\simeq 2$.

It thus appears reasonable that a significant fraction of the
passively-evolving subset of  $z\simeq 2-3$ galaxies with 10$^{11}$
M$_\odot$ assembled their mass in UV luminous star formation episodes
at higher redshifts.   While it is possible that these rest-UV
selected  galaxies would be detectable in deeper sub-mm observations,
or alternatively  that these galaxies may have undergone
short-duration sub-mm luminous  bursts of star formation in their
past, these results appear to indicate that  the ultra-luminous sub-mm
galaxies such as those presented in \cite{Capak08} and \cite{Daddi08}
do not necessarily provide the only route toward assembling  massive,
quiescent galaxies by $z\simeq 2$.  Constructing a complete  sample of
the precursors of this massive population of red galaxies must include
both rest-UV and sub-mm samples.

\section{Stellar Mass Densities at $z\simeq 4, 5, \rm{and} 6$}

We now estimate the stellar mass densities implied by our dropout
samples.  By integrating the mass functions presented in Figure 11, we
compute the stellar mass density brightward of the adopted rest-UV
magnitude limit (taken to be M$_{\rm{1500}}$=$-$20 for each of the
dropout samples).  This absolute magnitude limit corresponds to
apparent  magnitude limits of $i'_{775}\simeq 26.0$ for the B-drops,
z$_{850}\simeq 26.4$ for the V-drops,  and z$_{850}\simeq 26.7$ for
the $i'$-drops.  Using the $\tau$=100(300) Myr exponential decay
models, we obtain 1.1(1.4)$\times$10$^{7}$ M$_\odot$ Mpc$^{-3}$,
3.7(4.9)$\times$10$^6$ M$_\odot$ Mpc$^{-3}$, and
1.6(2.3)$\times$10$^{6}$ M$_\odot$  Mpc$^{-3}$ for the B, V, and
$i'$-drops, respectively (Figure 13).  Including all MIPS-detected
dropouts sources increases the derived mass densities to 1.6(1.9)
$\times$10$^{7}$ M$_\odot$ Mpc$^{-3}$ for the B-drops and
4.7(6.2)$\times$10$^6$ M$_\odot$ Mpc$^{-3}$ for the V-drops with  the
$\tau$=100(300) Myr star formation history.  The range of mass
densities quoted above are consistent with our previous measurements
at $z\simeq 5$ (5$\times$10$^{6}$ M$_\odot$ Mpc$^{-3}$,
\citealt{Stark07a}) and $z\simeq 6$ (2.5$\times$10$^{6}$ M$_\odot$
Mpc$^{-3}$, \citealt{Eyles07}). 

\begin{figure}
\figurenum{13} \epsscale{1.2}  \plotone{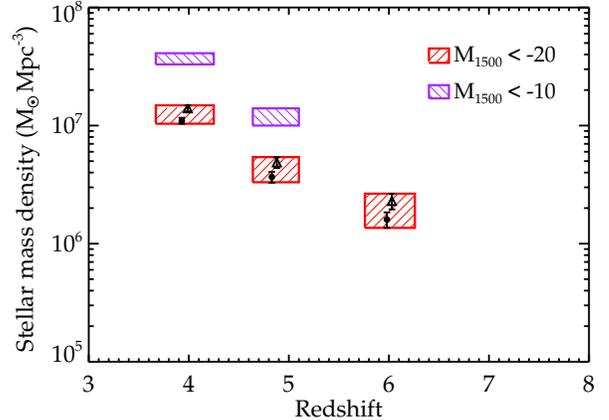} 
\caption{Stellar mass density of B, V, and $i'$-drops. The red
  rectangles  demarcate the stellar mass densities derived from our
  dropout sample with  an adopted magnitude limit of M$_{1500}=-20$.
  The circles(triangles) denote  mass densities and associated Poisson
  error  determined using CB07 models with an exponentially-declining
  star formation history with a decay factor of $\tau$=100(300) Myr.
  See  Figure 11 caption and \S5.1 for further discussion of the
  assumptions used  in deriving the stellar masses.  Purple squares
  demarcate rough estimates of the total  stellar mass density of
  galaxies brighter than M$_{1500}=-10$.  These values are determined
  by extrapolating the average relationship between stellar mass and
  M$_{1500}$ (Figure 9) to luminosities below the detection limits  of
  the GOODS dataset.  }
\end{figure}

As the total stellar mass density of galaxies at high redshift
provides a unique constraint on the past star formation history of the
universe,  it is of interest to estimate the contribution from
galaxies that are fainter  than the adopted rest-UV absolute magnitude
limit of M$_{1500}$=-20.  For the B and V-drop samples, this magnitude
limit is brighter than allowed by the S/N limits of the data (see
vertical lines in Figure 9).  If we compute the mass density of the
B-drop sample including all sources brighter than the 5$\sigma$ limit
of M$_{1500}\simeq -19.1$, we derive a mass density of
1.7(2.1)$\times$10$^{7}$ M$_\odot$  Mpc$^{-3}$.  Likewise, integrating
the stellar mass function of the V-drop sample including all sources
brighter than the 5$\sigma$ limit of M$_{1500}\simeq  -19.5$, we find
a mass density of 4.4(5.9)$\times$10$^6$ M$_\odot$ Mpc$^{-3}$.  Both
estimates exclude  the contribution of MIPS-detected sources.

In order to constrain the mass density of the population even fainter
than these limits, we need to know the average stellar mass to UV
luminosity ratio for galaxies below our detection thresholds.  We can
estimate this by extrapolating the trends present in the
M$_\star$-M$_{1500}$  relation of Figure 9.  Fitting a linear  model
to the observed trend for  our B-drop sample, we find the following
relation: $\overline{\rm{log_{10}  M_\star}}=-0.26 -
0.47(\rm{M_{1500}})$.  By extrapolating this function to  fainter
magnitudes and integrating the observed UV luminosity function
\citep{Bouwens07} to determine the number density of galaxies at each
M$_{1500}$, we can compute a rough estimate the contribution of {\it
  typical} faint UV sources to the total stellar mass density.  

Following this approach, we find that the entire population of typical
faint B-drops (e.g., those with average stellar masses in the
magnitude range $-19\lsim \rm{M_{1500}} \lsim -10$) comprises $\simeq
49$\% of the total stellar mass density.  This increases the total
stellar mass density in B-drops to  $\simeq
\rm{3.3(4.1)}\times$10$^{7}$ M$_\odot$ Mpc$^{-3}$ for the
$\tau$=100(300) Myr exponential decay models.  Applying the same
procedure to the  V-drops (assuming the same relationship between
stellar mass and M$_{1500}$ holds at $z\simeq 5$), we derive total
stellar mass densities of 1.0(1.4)$\times$10$^{7}$ M$_\odot$
Mpc$^{-3}$ for sources brighter than M$_{1500}$=-10.  We refrain from
making a similar extrapolation for the $i'$-drops given  the larger
uncertainties in this population's M$_\star$-M$_{1500}$  relation.

\section{Summary}
\label{sec:summary}

We have studied the growth of galaxies over $4\lsim z\lsim 6$ with the
goal of improving our understanding of galaxy evolution in the first 2
billion years of cosmic history.  Using the deep HST data available in
the two GOODS fields, we compiled samples of 2819 B-dropouts,  615
V-dropouts, and 166 $i'$-dropouts. The superior quality of the
multiwavelength data in GOODS (including optical data from HST
spanning four filters, near-infrared data  from the VLT and Subaru,
and  mid-infrared data from Spitzer) greatly helps  identify
foreground galaxies and stars which lurk in Lyman break dropout
samples.  Great care was taken to identify and remove these objects
from our sample, using HST morphologies to excise point sources likely
to be bright stellar contaminants and broadband SEDs spanning from the
optical to the infrared to remove low-redshift galaxies.  

Our final high-redshift sample consists of 2443 B-dropouts, 506
V-dropouts, and 137 $i'$-dropouts.  From this sample of galaxies, we
constructed sub-samples  of dropouts which are sufficiently unblended
with nearby foreground  objects in the deep Spitzer data to allow
accurate photometry.  In  general, this corresponds to $\gsim 35$\% of
the total samples.   We used the CB07 population synthesis models to
estimate stellar masses  and ages for this ``Spitzer clean'' subset.
We compared the inferred  properties to those derived from BC03
models, finding general agreement for solar metallicity models when
the redshift is above 5.  Below this redshift, the mid-IR data begin
to  probe the rest-frame near-infrared (which is affected by TP-AGB
stars), and the CB07 models therefore return masses which are
typically   10\% lower than those from BC03 models.  

Using this large database of stellar masses and rest-UV properties of
B, V, and $i'$-dropouts, we examine the evolution of star-forming
galaxies between $z\simeq 4$ and 6.  We summarize our main results
below.

1.   We find that the typical stellar masses and ages  of galaxies of
a fixed UV luminosity (uncorrected for dust extinction) do not evolve
strongly between $z\simeq 6$, 5,  and 4.  This argues against the
notion that the B-drops  are made up of galaxies that have been
steadily growing at fixed UV luminosity since $z\simeq 6$.  Instead,
these data suggest that each successive epoch studied in this survey
(e.g., $z\simeq 4$, 5, and 6) is likely  dominated by galaxies  that
have only recently (within the last $\simeq 300$ Myr) emerged at their
present luminosity.  This is consistent with the observed decline  in
the UV luminosity function between $z\simeq 4$ and 6
\citep{Bouwens07}.

2.  The evolution in the M$_\star$-M$_{1500}$ relation and the UV
luminosity function is potentially difficult to reconcile if the
galaxies  observed at $z\gsim 4$ assembled their stars with a constant
star formation  rate.  In this scenario, the inferred ages of galaxies
at one epoch are sufficiently large to violate the required decline in
the UV LF and the lack of strong evolution in the
M$_\star$-M$_{1500}$.  However if lower metallicity templates are
adopted,  the ages are lowered and the discrepancy reduced.  Deeper
infrared photometry will be of significant benefit in improving
constraints  on the inferred stellar populations.  

We also considered a steady growth scenario in which star formation is
becoming steadily more vigorous with increasing cosmic time between
$z\simeq 6$ and $z\simeq 4$,  as predicted  in the simulations of
\cite{Finlator06}.  In this case, the inferred ages of the galaxies
remain large, but since their precursors are fainter in the past, they
don't violate the observed evolution in the UV LF and
M$_\star$-M$_{1500}$ relation.  The only problem that potentially
stands in the way of this model is that it is inconsistent with the
short duty cycles implied by measurements of the clustering of $z\gsim
4$ LBGs \citep{Lee08}.  Additional spectroscopy and new
multiwavelength imaging in independent fields are required to confirm
these clustering measurements.

3.  We argue that episodic star formation may provide the best fit to
the wide range of observations considered in this paper.  In this
picture,  each epoch is dominated by recently emerged sources, which
remain luminous for no more than $\simeq 500$ Myr on average.  If the
UV luminous periods become more intense with cosmic time between
$z\simeq 6$ and 4, then this scenario could explain the evolution of
the M$_\star$-M$_{1500}$  relation and the UV LF.  If this is the
dominant mode of star formation  at $z\gsim 4$, then there should be a
significant population of relatively massive, quiescent sources that
are in between star formation episodes.  Future efforts at $z\simeq 4$
should help clarify whether this is the case.

4. The stellar mass function of UV luminous sources grows
significantly from $z\approx 6$ to $z\approx 4$,  typically increasing
by more than a factor of six in number density over  the entire
stellar mass range considered.  There is some evidence that the growth
in the most massive LBGs (M$_\star \simeq \rm{10^{11\pm0.5}}$) is a
factor of three greater than in less massive systems, implying that
this redshift range likely corresponds to the beginning of massive
(10$^{11}$ M$_\odot$) galaxy formation era.  

5. The rapid growth in the assembly of massive galaxies at $z\gsim 4$
indicates that at least $\simeq 20$\% of the $z\simeq 2-3$ quiescent
massive galaxies assembled their stellar mass at $z\gsim 3.5$.
Fitting  the increase in number density of UV luminous, massive
galaxies between $z\simeq 6$ and 2, we estimate that it is feasible
that more than 50\% of quiescent $z\simeq  2-3$ massive galaxies were
assembled in higher redshift LBGs.  These results imply that the high
star formation rate, dust-obscured  sub-mm galaxies are not the only
route toward assembling quiescent 10$^{11}$ M$_\odot$ systems by
$z\simeq 2-3$. 

6. Using the CB07 models with solar metallicity and a exponentially
declining star formation history with a decay factor of
$\tau$=100(300)  Myr, we derive the stellar mass density of rest-UV
selected dropouts brighter than  M$_{1500}=-20$: $\rho_\star \simeq
1.1$(1.4)$\times$10$^{7}$ M$_\odot$ Mpc$^{-3}$ at $z\simeq 4$,
3.7(4.9)$\times$10$^{6}$ M$_\odot$ Mpc$^{-3}$ at $z\simeq 5$, and
1.6(2.3) $\times$10$^6$ M$_\odot$ Mpc$^{-3}$ at $z\simeq 6$.  These
measurements are certain to underpredict the total stellar mass
density  since they  do not include the contribution from sources that
either do not satisfy are color selection criteria or lie faintward of
the absolute magnitude limit.  By  extrapolating the
M$_\star$-M$_{1500}$ relation from Figure 9 to M$_{1500}$=-10 and
integrating the UV luminosity functions of \cite{Bouwens07}, we
estimate that the population of faint galaxies may increase the mass
density estimates quoted above by a factor of 3-5$\times$.

\subsection*{ACKNOWLEDGMENTS}

We thank the referee for a very helpful report and acknowledge useful 
conversations with Tommaso Treu, Niv Drory, and Masami Ouchi.   DPS
thanks the Department of Astrophysics at the University of Oxford for its
hospitality while much of this work was being conducted and
acknowledges financial support from the STFC. RSE acknowledges
financial support from the Royal Society.  This paper is based on
observations made with the NASA/ESA Hubble Space Telescope, obtained
from the Data Archive at the Space Telescope Science Institute, which
is operated by the Association of Universities for Research in
Astronomy, Inc., under NASA contract NAS 5-26555. The {\em HST/ACS}
observations are associated with proposals \#9425\,\&\,9583 (the GOODS
public imaging survey).

\newpage

\bibliography{journals_apj,mybib}

\end{document}